\documentclass[journal=jacsat,manuscript=article]{achemso}
\usepackage{graphicx}
\usepackage{amssymb,amsfonts,amsmath}
\usepackage{float}
\usepackage{dcolumn}
\usepackage{bm}
\usepackage{color}
\usepackage[sort&compress]{natbib}
\usepackage{pdfpages}

\newcommand{\Na}{Na$_{2-\delta}$Mo$_6$Se$_6$}
\newcommand{\M}{$M_2$Mo$_6$Se$_6$}
\newcommand{\Tl}{Tl$_2$Mo$_6$Se$_6$}
\newcommand{\In}{In$_2$Mo$_6$Se$_6$}

\author{Diane Ansermet{$^\star$}}
\author{Alexander P. Petrovi\'c{$^\star$}\footnotetext{$^\star$ These authors contributed equally to this work.}}
\email{appetrovic@ntu.edu.sg}
\author{Shikun He}
\affiliation{Division of Physics and Applied Physics, School of Physical and Mathematical Sciences, Nanyang Technological University, 637371 Singapore}
\author{Dmitri Chernyshov}
\affiliation{Swiss-Norwegian Beamline, European Synchrotron Radiation Facility, 6 rue Jules Horowitz, F-38043 Grenoble Cedex, France}
\author{Moritz Hoesch}
\affiliation{Diamond Light Source, Harwell Campus, Didcot OX11 0DE, Oxfordshire, United Kingdom}
\author{Diala Salloum}
\affiliation{Sciences Chimiques, CSM UMR CNRS 6226, Universit\'e de Rennes 1, Avenue du G\'en\'eral Leclerc, 35042 Rennes Cedex, France}
\alsoaffiliation{Faculty of Science III, Lebanese University, PO Box 826, Kobbeh-Tripoli, Lebanon}
\author{Patrick Gougeon}
\author{Michel Potel}
\affiliation{Sciences Chimiques, CSM UMR CNRS 6226, Universit\'e de Rennes 1, Avenue du G\'en\'eral Leclerc, 35042 Rennes Cedex, France}
\author{Lilia Boeri}
\affiliation{Institute for Theoretical and Computational Physics, TU Graz, Petersgasse 16, 8010 Graz, Austria}
\author{Ole Krogh Andersen}
\affiliation{Max Planck Institute for Solid State Research, Heisenbergstrasse 1, D-70569 Stuttgart, Germany}
\author{Christos Panagopoulos}
\email{christos@ntu.edu.sg}
\affiliation{Division of Physics and Applied Physics, School of Physical and Mathematical Sciences, Nanyang Technological University, 637371 Singapore}

\title{Reentrant Phase Coherence in\\Superconducting Nanowire Composites}

\begin{document}
\newpage

\begin{abstract}
The short coherence lengths characteristic of low-dimensional superconductors are associated with usefully high critical fields or temperatures.  Unfortunately, such materials are often sensitive to disorder and suffer from phase fluctuations in the superconducting order parameter which diverge with temperature \textit{T}, magnetic field \textit{H} or current \textit{I}.  We propose an approach to overcome synthesis and fluctuation problems: building superconductors from inhomogeneous composites of nanofilaments.  Macroscopic crystals of quasi-one-dimensional {\Na} featuring Na vacancy disorder ($\delta\approx$~0.2) are shown to behave as percolative networks of superconducting nanowires.  Long range order is established \textit{via} transverse coupling between individual one-dimensional filaments, yet phase coherence remains unstable to fluctuations and localization in the zero-(\textit{T,H,I}) limit.  However, a region of reentrant phase coherence develops upon raising (\textit{T,H,I}).  We attribute this phenomenon to an enhancement of the transverse coupling due to electron delocalization.  Our observations of reentrant phase coherence coincide with a peak in the Josephson energy $E_J$ at non-zero (\textit{T,H,I}), which we estimate using a simple analytical model for a disordered anisotropic superconductor.  {\Na} is therefore a blueprint for a future generation of nanofilamentary superconductors with inbuilt resilience to phase fluctuations at elevated (\textit{T,H,I}).
\end{abstract}

\maketitle


Composite nanofilamentary systems offer a unique environment in which to study the effects of dimensionality and phase fluctuations on the superconducting transition~\cite{Klemm1985}. Such materials may be described as ``quasi-one-dimensional'' (q1D), since they exist in bulk macroscopic form yet exhibit an intense uniaxial anisotropy in their physical properties.  Modulating the transverse (inter-filamentary) coupling in these materials is of fundamental interest due to an expected dimensional crossover, whose impact on the electronic properties remains unclear~\cite{Giamarchi2004}.  From a practical perspective, one can also imagine synthetic ``ropes'' of coupled one-dimensional (1D) nanowires as future superconducting cables~\cite{Mishra2013}.  Aside from the obvious morphological advantage, there are powerful incentives to develop functional superconductors from nanoscale q1D building blocks.  Firstly, orbital limiting is suppressed in q1D superconductors~\cite{Turkevich1979}, thus increasing the critical field $H_{c2}$ to the Pauli limit (and in principle, a q1D superconductor with triplet pairing would be completely immune to magnetic fields).  Secondly, various mechanisms for enhancing the critical temperature $T_c$ exist in superconducting nanostructures, including tuning the density of states through van Hove singularities, shape resonances~\cite{Shanenko2006a}, strain-induced renormalization of the electronic structure~\cite{Tian2015} and shell effects~\cite{Kresin2006}.  Coupled arrays of nanowires therefore represent an attractive and realistic route towards developing new functional superconductors, as well as exploring the role of dimensionality in correlated electron systems.

\begin{figure} 
\centering
\includegraphics [width=0.5\linewidth] {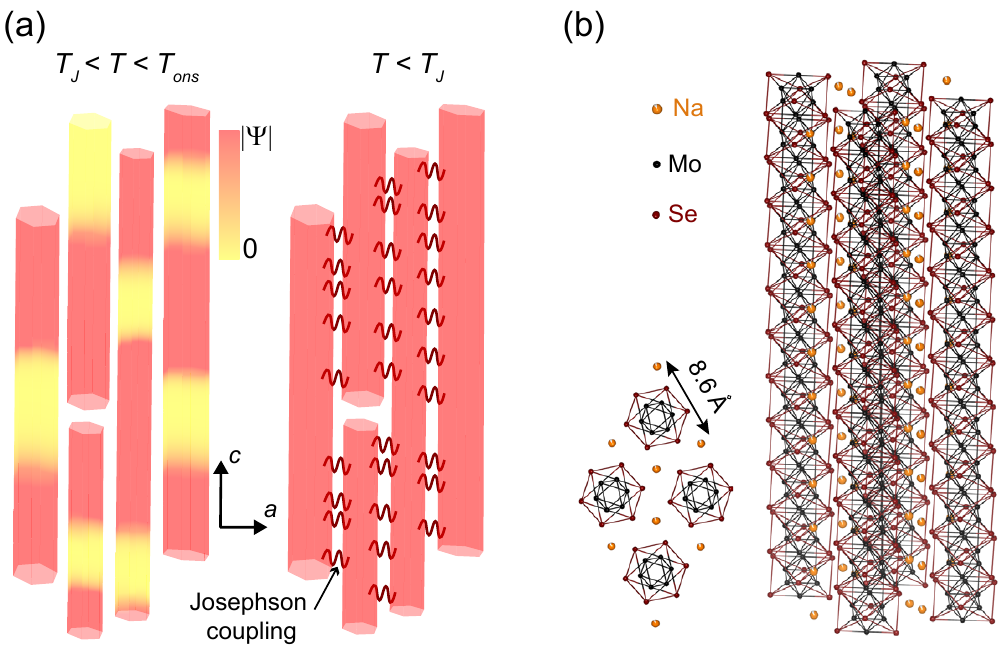}
\caption{\label{Fig1} {\Na}: an ideal q1D nanofilamentary superconductor.  (a) In a q1D superconductor, the intense uniaxial anisotropy leads to a 2-step superconducting transition in which pairing within individual superconducting filaments occurs prior to the establishment of global phase coherence.  Above $T_{J}$, the crystal behaves as an array of decoupled 1D filaments due to the electronic anisotropy.  For stoichiometric Na$_2$Mo$_6$Se$_6$, the filament diameter is 1 unit cell (u.c.); however variations in the Na vacancy distribution across macroscopic crystals could potentially create broader filaments several u.c. wide.  Neighboring filaments are phase incoherent and fluctuations suppress $|\Psi|$ to zero at certain points (yellow shading), preventing the establishment of a zero-resistance state.  Below $T_{J}$, phase coherence develops from transverse Josephson coupling between filaments and a dimensional crossover occurs from 1D to q3D superconductivity.  (b) Crystal structure of {\Na}, viewed parallel to the $c$ axis (left) and at an oblique angle to $c$ (right).  A clear structural parallel exists with the ideal q1D superconductor shown in (a).}  
\end{figure}

Although fabrication techniques for such nanofilamentary composites are in their infancy, we may hasten their development by studying quasi-one-dimensional (q1D) crystals featuring chain-like structures, which behave as weakly-coupled arrays of parallel nanowires.  Within the superconducting phase, coupling between nanowires is expected to occur \textit{via} the Josephson effect, \textit{i.e.} phase-coherent Cooper pair tunnelling~\cite{Schulz1983a}.  An identical process is responsible for establishing long-range order between crystal planes in highly two-dimensional (2D) superconductors, including cuprates~\cite{Kleiner1992} and pnictides~\cite{Moll2014}.  However, the superconducting transitions in q1D materials exhibit important differences compared with 2D or 3D superconductors, which we summarize in Fig.~\ref{Fig1}(a).  In a q1D system, transverse Josephson coupling (\textit{i.e.} phase-coherent inter-chain Cooper pair tunnelling) is a second order process~\cite{Giamarchi2004} which occurs below temperature $T_J\!\geq\!t_\perp^2/t_{/\!/}$, where $t_\perp$ and $t_{/\!/}$ are the single-particle hopping energies perpendicular and parallel to the structural chains.  For most q1D superconductors, $T_c~{\ll}~t_\perp^2/t_{/\!/}$ and there is a direct transition from the normal state to a quasi-three-dimensional (q3D) superconducting phase (a detailed discussion is provided in section I of the Supporting Information (SI)).  However, if the anisotropy and the pairing interaction are both sufficiently large, the onset temperature for superconducting fluctuations $T_{ons}$ may be greater than $t_\perp^2/t_{/\!/}$: in this case, a ``2-step'' superconducting transition occurs.  Below $T_{ons}$, 1D superconductivity develops within individual chains but topological defects (phase slips) locally suppress the amplitude of the superconducting order parameter $\left|\Psi\right|$ to zero, creating a resistive state identical to that of a single nanowire~\cite{Bezryadin2000,Arutyunov2008,Altomare2013}.  Subsequently, a 1D$\rightarrow$q3D dimensional crossover occurs within the superconducting state: phase coherence (and hence long range order) are established at $T_J<T_{ons}$.

The {\M} family~\cite{Potel1980} ($M$~=~Tl,~In,~Na,~K,~Rb,~Cs) are archetypal q1D materials, in which infinite-length (Mo$_6$Se$_6$)$_\infty$ chains are aligned along the $c$ axis of a hexagonal lattice (Fig.~\ref{Fig1}(b)).  $M$ ions are intercalated between the chains and act as a charge reservoir for a single 1D conduction band of predominant Mo $d_{xz}$ character.  Transverse coupling occurs \textit{via} the $M$ ions and the electronic anisotropy may hence be tuned by selection of $M$.  Furthermore, {\M} crystals typically display small $M$ ion deficiencies~\cite{Brusetti1994a}, constituting an intrinsic disorder.  These $M$ vacancies reduce the coupling between (Mo$_6$Se$_6$)$_\infty$ chains and break them electronically into finite-length segments.  {\Tl} and {\In} are already known to exhibit superconducting ground states~\cite{Armici1980a,Petrovic2010,Bergk2011}; {\M} are therefore ideal materials in which to investigate the behavior of future superconducting composites made from coupled nanowire arrays.

In this work, we show that disordered single crystals of superconducting {\Na} display ``reentrant'' characteristics, where phase coherence is stabilized by an increased inter-filamentary coupling within a region of non-zero $(T,H,I)$ phase space.  The ability to control the transverse phase coherence by modulating $(T,H,I)$ arises from the sensitivity of the inter-filamentary coupling to electron localization, which is gradually suppressed as $(T,H,I)$ increase.  Reentrant phase coherence is a highly desirable property, since maintaining phase stiffness~\cite{Emery1994} at elevated $(T,H,I)$ is perhaps the greatest challenge to the development of new functional superconductors~\cite{Gurevich2011a}.  Our results pave the way towards synthesizing nanofilamentary materials whose superconducting properties are enhanced rather than destroyed within the high $(T,H,I)$ domain.

\section{RESULTS}
\subsection{Electronic and crystal structure of \Na}  

Our identification of {\Na} as a model nanofilamentary superconductor was motivated by a combination of electronic structure calculations and crystal growth considerations.  Firstly, strong crystalline anisotropy (\textit{i.e.} a weak transverse coupling) is an essential prerequisite for accurately simulating a nanofilamentary superconductor.  Using \textit{ab initio} density functional theory, we have calculated transverse Josephson coupling temperatures $t_\perp^2/t_{/\!/}=$~4.4~K, 3.0~K, 1.0~K for {\Tl}, {\In} and Na$_2$Mo$_6$Se$_6$ respectively (see SI section I for details).  The ground state of {\Na} has until now remained unknown; however the extremely weak transverse coupling implies that any superconducting order parameter in this material will be highly anisotropic and exhibit a low phase stiffness.  Secondly, the small Na atomic radius and elevated growth temperature (1750~$^{\circ}$C) are expected to increase the Na ion mobility during crystal synthesis, resulting in a substantially larger Na deficiency than the usual $\sim2.5-5$\% observed in {\Tl}~\cite{Brusetti1994a}.  Increasing the Na vacancy concentration (and hence the disorder) will result in crystals which are more realistic analogies to an inhomogeneous nanofilamentary array.

We therefore synthesized a series of needle-like {\Na} single crystals with typical lengths 2-3~mm (see Methods for details).  Synchrotron X-ray diffraction measurements indicate a typical 10\% Na deficiency, \textit{i.e.} $\delta\approx$~0.2, although the (Mo$_6$Se$_6$)$_\infty$ crystal superstructure remains highly ordered.  [A complete structural refinement is included in the Supporting Information.]  The influence of the Na vacancy-induced disorder can clearly be seen in the electrical resistivity $R(T,H,I)$, which rises due to localization at low temperature (Fig.~S6) before the crystals undergo a transition to a superconducting ground state (Figs.~\ref{Fig2},\ref{Fig3}).  Here we focus on the extent and control of phase-coherent superconductivity as a function of $(T,H,I)$ alone.  A brief discussion of the possibility of tuning the superconducting ground state by varying the disorder level may be found in SI section VIII.

\subsection{Dimensional crossover in the superconducting transition}  
We first demonstrate that {\Na} displays the 2-step transition outlined in Fig.~\ref{Fig1}(a).  Figure~\ref{Fig2} shows the superconducting transitions in $R(T)$ for a typical {\Na} single crystal with $T_{ons}=$~2.7~K.  Na vacancy disorder will reduce the coherence length $\xi$ and hence the energy barrier to topological defect formation~\cite{Arutyunov2008,Altomare2013} in the 1D regime ($T_{J}<T<T_{ons}$).  We therefore anticipate an important contribution to $R(T)$ from thermally-activated phase slips (TAPS) along individual (Mo$_6$Se$_6$)$_\infty$ nanowires within this temperature range.  To model our data, we adapt the well-known Langer-Ambegaokar-McCumber-Halperin (LAMH) TAPS model to describe an array of superconducting nanowires and proceed to fit the $R(T)$ superconducting transitions (Fig.~\ref{Fig2}).  A detailed discussion of the LAMH model may be found in SI section IIIB,C.  Directly below $T_{ons}$, our model accurately reproduces $R(T)$ independently of the excitation $I$: this indicates a universal onset of fluctuating 1D superconductivity.  To the best of our knowledge, {\Na} is the first bulk q1D superconductor to be accurately described by any 1D phase slip theory.

However, 1D LAMH theory can only describe our data over a finite temperature range $\sim$~0.4~K.  As the temperature is reduced further, an anomaly appears in each $R(T)$ curve whose position is displaced to lower temperature as $I$ increases.  For $I\leq0.1$~mA, the LAMH regime in $R(T)$ is terminated by a sharp peak: this corresponds to a suppression of single-particle tunnelling between phase-incoherent superconducting filaments, followed by the onset of phase coherence (\textit{i.e.} inter-chain Cooper pair tunnelling) at lower temperature.  $R(T)$ subsequently forms a finite-resistance plateau instead of falling to zero, which we attribute to isolated barriers such as micro-cracks and twin boundaries separating macroscopic phase-coherent superconducting regions (SI sections IIID, VIII).  Eventually $R(T)$ rises again as $T\rightarrow0$: this is the first experimental signature of reentrance.  In contrast, for $I>0.1$~mA the resistance begins to diverge from the LAMH model around $T\sim1.8$~K but continues to fall without forming a plateau, and eventually saturates with no upturn in the $T\rightarrow0$ limit.  The sharp peak is smeared into a broad hump (Fig.~S3), which we attribute to pair-breaking effects from the increased current.  Together, these features indicate the emergence of a phase-coherent q3D superconducting state composed of coupled 1D filaments.

Our electronic structure calculations predict that a 1D$\rightarrow$q3D dimensional crossover for two-particle hopping (\textit{i.e.} Josephson coupling) should occur at temperature $t_\perp^2/t_{/\!/}=$~1.0~K (SI Section I).  However, the anomaly in the $R(T)$ data and the deviation from LAMH fits suggest that transverse coupling develops at higher temperature 1.4~K~$\lesssim T_J \lesssim$~2.0~K. This increase in $T_J$ relative to our theoretical expectations may be attributed to two factors: firstly, any defects (including Na vacancies) strongly reduce the effective $t_{/\!/}$ due to the ease of blocking electron motion along a single (Mo$_6$Se$_6$)$_\infty$ filament.  Although it may initially seem counter-intuitive for inter-chain defects to influence intra-chain transport, each Na vacancy not only removes one electron from the conduction band (which is predominantly of Mo $d_{xz}$ character), but also locally modifies the crystal field.  In such intensely anisotropic materials, even minor crystal field inhomogeneities can lead to an enhanced back-scattering at low temperature~\cite{Fisher1997a}.  Secondly, $T_J$ is believed to be enhanced to higher temperatures by increasingly strong electron-electron interactions~\cite{Giamarchi2004}, although the behavior of q1D electron liquids below the single-particle dimensional crossover temperature ($T_x \leq t_\perp\sim$~120~K in {\Na}) remains to be completely understood.

Interestingly, our experimental $R(T)$ and voltage-current $V(I)$ data share all the features of the well-known Berezinskii-Kosterlitz-Thouless (BKT) transition, which establishes long-range order in 2D materials.  This suggests that an exponential divergence in the transverse phase correlation length occurs close to $T_J$ in q1D materials.  The similarity between 2D systems exhibiting BKT transitions and q1D superconductors becomes apparent upon considering the phase of the order parameter on each 1D filament (Fig.~S4(a)).  In the plane perpendicular to the filaments, the spatial variation of the phase satisfies 2D~$XY$ symmetry.  However, the validity of BKT physics in q1D materials is neither obvious nor trivial, since it would imply that phase fluctuations parallel to the filaments do not influence the onset of transverse phase coherence.  Nevertheless, a BKT-type analysis (SI section IIIE) of our transport data yields $T_J=$~1.71~K, in good agreement with the anomalies which we observe in our $R(T)$ data.  

Combining our LAMH fitting parameters and $T_J$ enables us to estimate a typical filament diameter $\sim$~0.4-0.7~nm (SI section IV).  This corresponds closely to the (Mo$_6$Se$_6$)$_\infty$ chain diameter of 0.60~nm and the hexagonal lattice parameter $a=$~0.86~nm, indicating that single (Mo$_6$Se$_6$)$_\infty$ chains behave as 1D superconducting nanowires.  Our analysis also indicates that current flows inhomogeneously through {\Na} and is supported by simulations of a disordered q1D conductor with an anisotropic random resistor network (SI section V).  We attribute the inhomogeneous flow to the enhanced influence of disorder in 1D materials: above $T_J$, defects (e.g. Na vacancies) restrict transport along individual (Mo$_6$Se$_6$)$_\infty$ chains, forcing the current to follow a highly percolative route.

\subsection{Reentrant phase coherence}  
Aside from the 2-step 1D~$\rightarrow$~q3D transition, the major feature of interest in Fig.~\ref{Fig2} is the rise in resistance as $T\!\rightarrow\!0$ for $I\leq0.1$~mA.  This reversion to a fluctuation-dominated state suggests that the transverse phase coherence is fragile and reentrant.  We track the evolution of the reentrance with current in Fig.~\ref{Fig3}(a), where three important trends may be identified.  Firstly, the superconducting transition is conventionally suppressed to lower temperature as $I$ increases.  However, the temperature dependence of the critical current $I_c(T)$ does not follow the standard Bardeen relation derived for bulk superconductors~\cite{Bardeen1962} (Fig.~\ref{Fig3}(a) inset), remaining unusually large at high temperature.  Secondly, for $T<1$~K the resistance falls as $I$ rises.  This indicates that elevated currents induce reentrant phase coherence even in the $T\rightarrow0$ limit.  Thirdly, the resistance rises (signalling a loss of phase coherence) upon reducing the temperature for $I\leq0.1$~mA.  Long range superconducting order is therefore only stable within a well-defined region of non-zero $(T,I)$ phase space.

To determine whether the phase coherence is also reentrant in magnetic fields, we measure the magnetotransport perpendicular and parallel to the $c$ axis (Fig.~\ref{Fig3}(b-e)).  A clear dichotomy is observed between data acquired using low (Fig.~\ref{Fig3}(b,d)) and high (Fig.~\ref{Fig3}(c,e)) currents.  The $R(T)$ transitions in $I_{\mathrm{high}}\equiv$~0.6~mA resemble those of a conventional superconductor (albeit substantially broadened) and no trace of reentrance is visible.  In this case, transverse phase coherence has already been stabilized by the large current and hence the magnetic field exhibits a purely destructive effect on superconductivity.  In contrast, for $I_{\mathrm{low}}\equiv$~1~$\mu$A, $R(T)$ rises at low temperature, indicating field-induced reentrance.  For $H_\perp=$~0.75~T, $R(T)$ falls again below $T=$~0.5~K, indicating double-reentrant behavior~\cite{VanderZantHS1996,Hadacek2004a}.  Crucially, both phase coherence and double-reentrance are absent as $T\rightarrow0$ in our zero-field data acquired at low current.   This implies that a sequence of superconducting fluctuations, suppression of quasiparticle tunnelling and eventual macroscopic phase coherence (which has been suggested to create ``double dips'' in $R(T\rightarrow0)$ in other inhomogeneous superconductors~\cite{Gerber1990a,Parendo2007}) cannot be responsible for these data.  Instead, the double-reentrance may be a signature of a divergent $H_{c2}$ caused by mesoscopic fluctuations~\cite{Spivak1995}.

Regardless of the applied current, the magnetotransport varies strongly with the field orientation, as expected for a q1D superconductor.  We quantify this anisotropy \textit{via} the temperature dependence of the upper critical fields $H_{c2\perp,/\!/}(T)$, which we plot in Fig.~\ref{Fig3}(f).  Using the Werthamer-Helfand-Hohenberg (WHH) model to estimate $H_{c2/\!/}(T\!=\!0,I_\mathrm{small})$ and subsequently applying anisotropic GL theory (SI Section VI), we extract coherence lengths 14.4~nm~$\leq\xi_{/\!/}(0)\leq$~21.0~nm, 4.28~nm~$\leq\xi_\perp(0)\leq$~4.58~nm and an anisotropy $\epsilon\equiv\xi_{/\!/}/\xi_\perp$ of 3.14~$\leq\epsilon\leq$~4.90.  This value is lower than the 12.6 reported for {\Tl}~\cite{Petrovic2010}, in spite of the increased electronic anisotropy: $t_{/\!/}/t_\perp=$~86 for {\Na}, \textit{versus} 31 for {\Tl}. Two factors are responsible for this: firstly, disorder from the high Na vacancy density strongly suppresses $\xi_{/\!/}$ and hence $\epsilon$.  This is exposed by the orbitally-limited values for $H_{c2\perp}$ in {\Tl} and {\In}: 0.47~T and 0.25~T respectively~\cite{Petrovic2010}, an order of magnitude lower than the 3.7-5~T measured for {\Na}.  [Note that this reduction in $\epsilon$ also supports the previously-discussed enhancement of $T_J$ above $t_\perp^2/t_{/\!/}\equiv$~1.0~K.]  Secondly, our estimated $H_{c2/\!/}\approx$~16-18~T exceeds the weak-coupling BCS Pauli limit $H_P\equiv1.84T_{ons}=$~5.0~T.  This means that paramagnetic rather than orbital limiting is likely to suppress superconductivity for $H/\!/c$ and hence GL theory may only provide an upper limit for $\xi_\perp$.

A striking divergence of $R(T)$ is observed at low current for $T\lesssim0.8$~K and $H_{\perp}\gtrsim2.5$~T (Fig.~\ref{Fig3}(d)).  This is a signature of magnetic field-induced Anderson localization, predicted to occur in q1D materials when a field is applied perpendicular to the high-conductivity axis~\cite{Dupuis1992b}.  In the normal state, the conditions for localization are $\hbar\omega_c \gg t_\perp$ and $k_BT \ll t_\perp$ (where $\omega_c$ is the cyclotron frequency $\mu_0He/m_e$).  Since electrons are paired for $T<T_{ons}$, Josephson tunnelling supplants single-particle hopping as the transverse coupling mechanism and we replace $t_\perp$ in the above conditions with $k_BT_J$.  Cooper pair localization is therefore expected for $H_\perp\gg$~1.3~T and $T\ll$~1.7~K, in good agreement with our data.  The resistance does not diverge for high currents (Fig.~\ref{Fig3}(e)) due to the pair-breaking effect of the increased current: in this case, field-induced localization is only expected for $\mu_0H>t_{\perp}m_e/\hbar{e}\equiv90$~T.

\subsection{Experimental phase diagram}  
The reentrance visible within our $R(T)$ data in Figs.~\ref{Fig2},\ref{Fig3} may be summarized by independently scanning $R(T)$, $R(H_{\perp,/\!/})$ and $R(I)$ (Fig.~\ref{Fig4}(a)).  In the superconducting phase of {\Na}, $R$ is always minimized at non-zero $(T,H,I)$: this is in direct contrast to the behavior of a conventional bulk superconductor, where phase fluctuations in $R$ are invariably minimized as $(T,H,I)\rightarrow0$.  We highlight the fact that the $R(H_{/\!/,\perp})$ and $R(I)$ curves were acquired at $T=0.1$~K, thus confirming that transverse phase coherence is reentrant even as $T\rightarrow0$ for sufficiently large magnetic fields or currents.  We also note that the magnetoresistance (MR) is initially positive for $R(H_{\perp,/\!/})$, before falling steeply to its minimum value as transverse coupling is established.  Conversely, quasiparticle tunnelling contributions to the transport would be expected to yield a gradual, monotonic negative MR prior to reentrance at low temperature~\cite{Kunchur1987}.  The absence of such a feature from our data provides further evidence that quasiparticle tunnelling does not play a major role in the reentrant behavior of {\Na}.

We map the extent of phase coherence in {\Na} by assembling our experimental data into a single phase diagram (Fig.~\ref{Fig4}(b)).  $H_{c2}(T)$ and $I_c(T)$ (circles) from Figs.~2,3 accurately describe the evolution of the superconducting transition, but do not capture the loss of phase coherence at lower temperature.  To track this loss of coherence, we therefore define a ``reentrance threshold'' temperature $T_R(H,I)$ (stars) using the minima in $R(T,H)$ from Figs.~3,4(a).  $T_R$ varies from zero to 1.6~K, illustrating how the reentrant regime spans a broad temperature range as $(H,I)$ are tuned.  At temperatures below $T_R(H,I)$, neighboring filaments are phase-incoherent and {\Na} exhibits the finite resistance characteristic of a fluctuating 1D superconductor.  The volume enclosed by $H_{c2}(T)$, $I_c(T)$ and $T_R(H,I)$ in $(T,H,I)$ phase space hence contains a reentrant ``shell'' of phase-coherent superconductivity (dark red shading in Fig.~\ref{Fig4}(b)), which surrounds a phase-incoherent regime as $(T,H,I)\rightarrow0$ (pink shading in Fig.~\ref{Fig4}(b)).  Even if two of $(T,H,I)$ fall to zero, reentrance can still occur if the third parameter is sufficiently large: for example, phase coherence is still reentrant as $(T,H)\rightarrow0$ above a threshold $I\sim0.1$~mA.

\subsection{Mechanisms for reentrance} 
In an inhomogeneous superconductor, reentrance may occur if the Josephson energy $E_J(T,H,I)$ rises with respect to the thermal energy $k_BT$ (or the Coulomb energy $E_C$ in granular materials)~\cite{VanderZantHS1996,Simanek1979,Efetov1980,Simanek1985,Zwerger1987,Belevtsev1989a}.  $E_J$ is a measure of the phase stiffness (and hence the coupling strength) between neighboring superconducting regions: microscopically, $E_J$ is proportional to the spatial overlap of the Cooper pair wavefunctions from each region.  A rise in $E_J$ increases the energy cost of creating phase discrepancies, hence facilitating Cooper pair (Josephson) tunnelling between the regions.  Once $E_J$ exceeds a threshold value of the order of $k_BT+E_C$, global phase coherence is established.  Previous experimental observations of reentrance attributed to Josephson effects have generally occurred in amorphous, ultra-thin or granular films~\cite{Hadacek2004a,Parendo2007, Kunchur1987,Belevtsev1989a,Lin1984,Orr1985,Belevtsev1987,Jaeger1989,Hollen2011,Heera2013a}.

{\Na} is not an inhomogeneous or granular superconductor in the traditional sense, where phase coherence is determined by the ratio of $E_J$ to $E_C$ for individual grains~\cite{Efetov1980,Simanek1985}.  Instead, it is a crystalline superconductor whose normal state is a localized metal, due to the combination of Na vacancy disorder and intense 1D anisotropy.  Signatures of localization are clearly visible in the normal-state resistance $R_{\mathrm{NS}}$, which diverges as $T\!\rightarrow\!0$ and displays a large negative magnetoresistance (Fig.~S6).  Localization causes electronic states which lie close in energy to become widely separated in space~\cite{Sadovskii1997}, leading to a characteristic activation energy~\cite{Shklovskii1984} $E_a(T,H)$.  Although we deduce the presence of localization from the normal-state transport, its influence persists within the superconducting phase, where pairing occurs between localized electrons (provided that $\xi_{/\!/}$ remains shorter than the localization length~\cite{Sadovskii1997}).  As the temperature falls, the electron wavefunctions become increasingly localized and the inter-filamentary pair hopping energy begins to fall below the $t_\perp^2/t_{/\!/}$ limit imposed by the electronic anisotropy.  This implies a progressive reduction in the wavefunction overlap between neighbouring filaments: localization is suppressing the transverse coupling and hence $E_J$.  Eventually, $E_J(T,H,I)$ falls below the threshold for phase coherence ${\sim}k_BT$ at $T = T_R(H,I)$.  The influence of localization is accentuated by an emergent spatial inhomogeneity in the superconducting order parameter~\cite{Dubi2007,Sacepe2008,Lin2012b} which may locally suppress pairing, leading to further reductions in $E_J$.  To achieve reentrant phase coherence, it is necessary to increase $E_J$ by delocalising the electrons.  In principle, this may be achieved by thermal activation (raising $T$), reducing the barrier height between localized states (raising $I$) or Zeeman-splitting localized energy levels~\cite{Fukuyama1979} (raising $H$).  

Let us now attempt to model the effects of localization on the transverse phase coherence.  The $(T,H,I)$ evolution of the inter-filamentary pair hopping energy cannot be determined experimentally: even if the transverse resistance $R_\perp(T,H,I)$ could be accurately measured (an extremely challenging task due to the crystal morphology, anisotropy and disorder), it would contain inseparable contributions from single-particle and pair hopping, and fall to zero (depriving us of information) upon establishing phase coherence.  Instead, we estimate the $(T,H,I)$ dependence of $E_J$ within an analytical framework originally proposed by Belevtsev \textit{et al.} for inhomogeneous superconductors~\cite{Belevtsev1989a}:
\begin{equation}\label{eq1}
E_J \propto \frac{1}{R_T}\triangle\tanh\left(\frac{\triangle}{2k_BT}\right)
\end{equation}
where $R_T$ is the resistance between superconducting filaments and $\triangle$ is the pairing energy.  For reentrant superconductivity, a peak is expected to form in $E_J$ at finite $(T,H,I)$.  Since $\triangle\tanh\left(\triangle/2k_BT\right)$ falls monotonically to zero as $(T,H,I)$ increase, a peak in $E_J$ at non-zero $(T,H,I)$ can only develop if the drop in $\triangle\tanh\left(\triangle/2k_BT\right)$ is initially compensated by a larger reduction in $R_T$ (see Fig.~S7).  In the granular superconductors for which equation~(\ref{eq1}) was originally derived, the origins of this reduction are well understood.  Energy levels in individual grains are quantized, creating an activation energy for electron transfer: $R_T$ therefore falls exponentially as $T$ rises.  Increasing $I$ also reduces $R_T$, since the associated increase in voltage diminishes the effective barrier height between grains.  If $R_T$ furthermore exhibits negative magnetoresistance, then reentrance may occur upon raising $T$, $H$ or $I$.  Applying a similar scenario in {\Na}, we extract $E_a(T,H)$ from $R_{\mathrm{NS}}(T,H)$ and treat this term analogously to the granular activation energy discussed above.  We do not consider any Coulomb contribution to the reentrance, since {\Na} is crystalline and  $R_{\mathrm{NS}}(T)$ does not obey the Efros-Shklovskii hopping law~\cite{Shklovskii1984} indicative of strong Coulomb repulsion.  Using a similar current dependence for the inter-filamentary electron transfer rate to that in granular superconductors (see SI Section VII), we may then utilize the framework of equation~\ref{eq1} to estimate the evolution of $E_J(T,H,I)$.

Although we cannot calculate absolute values of $E_J$ (since several scaling parameters remain unknown), we may nevertheless establish the existence and location of any peaks in $E_J(T,H,I)$.  The resultant curves are plotted above the corresponding $R(T,H,I)$ data in Fig.~\ref{Fig4}(a).  Independently of changing $T$, $H$ or $I$, a peak appears in $E_J$.  The peak positions approximately correspond to the resistance minima and hence the reentrance threshold $T_R(H,I)$.  An exception to this trend occurs for $H/\!/c$, where the minimum in $R(H_{/\!/})$ occurs at a lower field than the peak in $E_J(H_{/\!/})$.  This may be caused by the WHH model overestimating the true $H_{c2/\!/}$ for {\Na} due to paramagnetic limiting.  Finally, we simulate our experimental phase diagram, plotting the theoretical $\triangle(T,H,I)$ instead of $H_{c2}(T),I_c(T)$ and calculating $T_R(H,I)$ by evaluating the locus of the peaks in $E_J$ within the $(H,T)$ and $(I,T)$ planes.  The results are shown as an inset to Fig.~\ref{Fig4}(b): a clear agreement is visible between our experimental and simulated phase diagrams.

\section{DISCUSSION}
Our data indicate that phase coherence in superconducting {\Na} is stabilized by a large reentrant coupling between electron-doped (Mo$_6$Se$_6$)$_\infty$ chains.  The reentrance is a direct consequence of electron localization induced by Na vacancy disorder.  To clarify this mechanism, we sketch an inhomogeneous q1D superconductor composed of finite-length dirty nanofilaments in Fig.~\ref{Fig5}(a).  The conduction band electrons become localized at low temperature, leading to a rise in $R(T)$ and a negative MR (Fig.~\ref{Fig5}(b)).  In the $(T,H,I)\rightarrow0$ limit, Josephson coupling between the nanofilaments is suppressed, resulting in phase-incoherent fluctuating superconductivity.  Here we must point out a flaw in our model, which predicts $R_T\rightarrow\infty$ as $T\rightarrow0$, independently of $(H,I)$.  This would imply a vanishing Josephson energy $E_J\rightarrow0$ and an invariable loss of phase coherence as $T\rightarrow0$.  In contrast, our experiments suggest that phase coherence remains stable at elevated $(H,I)$, even at $T=0$ (Fig.~\ref{Fig4}(b)).  $R_T$ must therefore remain finite (\textit{i.e.} metallic) at $T=0$.  

This discrepancy between data and model is linked to the nature of the disorder-induced superconductor-insulator transition in {\Na}.  It is possible that the typical disorder level in our crystals is sub-critical, \textit{i.e.} $R(T\!=\!0)\ll\infty$ and the q1D hopping model which reproduces our experimental $R(T)$ data (Fig. S6, SI Section VII) is merely valid over a finite temperature range.  The superconductor-insulator transition may also be replaced by a superconductor-metal-insulator transition for sufficiently large $(H,I)$, as suggested by the finite resistance which we measure as $T\rightarrow0$ in Fig.~\ref{Fig3}(e).  We note that zero-temperature metallic states have been predicted~\cite{Feigelman2001} and observed~\cite{Eley2011,Lin2012b} in Josephson-coupled superconducting arrays as well as amorphous Nb$_x$Si$_{1-x}$~\cite{Humbert2014}.  However, the physical concept which underlies our model (\textit{i.e.} the formation of peaks in $E_J$ at non-zero $(T,H,I)$) remains valid regardless of the zero-temperature state of {\Na}: equation~(\ref{eq1}) continues to yield values for $T_R(H,I)$ in good agreement with our experimental data, even as our calculated $E_J$ values become vanishingly small in the $T\rightarrow0$ limit.

Upon increasing $(T,H,I)$, the electrons are delocalized due to thermal activation, Zeeman level splitting and reduced barrier heights.  Cooper pairs begin to tunnel between the (Mo$_6$Se$_6$)$_\infty$ filaments and phase coherence is initially stabilized rather than destroyed (Fig.~\ref{Fig5}(c)).  Provided that the order parameter exhibits $s$-wave symmetry, we conclude that disorder in a nanofilamentary composite is beneficial to superconductivity, shortening $\xi_{/\!/}$ (thus raising $H_{c2\perp}$) while facilitating reentrant phase coherence.  Although we acknowledge that $T_{ons}$ is low ($\lesssim3$~K) in {\Na} (due to the combination of a small density of states at the Fermi level and weak electron-phonon coupling), the electron delocalization mechanism responsible for reentrance remains active at temperatures at least an order of magnitude higher (see Fig.~S6).  We stress that in the presence of a pairing interaction, there is no obvious thermal limitation to this reentrance mechanism: negative MR and $dR/dT<0$ can both persist up to room temperature in disordered nanomaterials~\cite{Wang2006}.  Furthermore, we anticipate that competing instabilities (such as density waves partially gapping the Fermi surface) which generate similar normal-state transport properties in other q1D superconductors may also enable reentrant phenomena to develop.

In addition to the transverse coupling detailed above, we cannot rule out some contribution from intra-filamentary defects - \textit{i.e.} Josephson coupling across barriers cutting (Mo$_6$Se$_6$)$_\infty$ chains - to the observed reentrance in {\Na}.  Such defects are certain to be present in our samples, and we believe that they share responsibility for the finite-resistance plateaus which form in $R(T)$ at low current (Figs.~\ref{Fig2}, S8; SI Section VIII).  However, supercurrents can percolate around such barriers without large resistive losses, provided that the chains are phase-coherent: this explains why $R(T)$ remains small and approximately constant in the plateau region.  Since phase coherence is established at $T_J>T_R$, the clear rise in the resistance for $T<T_R(H,I)$ must correspond to the loss of transverse phase coherence.  This is an inter- rather than intra-filamentary effect.  The key role of transverse coupling in establishing phase coherence is confirmed by Fig.~\ref{Fig3}(d), in which 1D localization of Cooper pairs completely suppresses transverse coupling between (Mo$_6$Se$_6$)$_\infty$ chains for sufficiently large transverse magnetic fields and low temperatures.  If the transverse coupling were not the principal factor controlling the resistance below $T_{J}$, $R(T)$ would not diverge as $T\rightarrow0$, in direct contrast with our data.    

In summary, we have demonstrated that {\Na} single crystals behave as ideal inhomogeneous nanofilamentary superconductors, in which a 1D$\rightarrow$q3D dimensional crossover occurs \textit{via} transverse Josephson coupling.  Inhomogeneity and disorder result in electron localization, which is evident from the normal-state magnetotransport: the superconducting nanofilaments consequently become decoupled at low temperatures.  However, transverse phase coherence is reentrant upon increasing $(T,H,I)$, since the electrons are progressively delocalized and hence the Cooper pair wavefunction overlap rises between neighbouring filaments.  This reentrance constitutes a key advantage over homogeneous materials, whose superconducting properties generally deteriorate at elevated $(T,H,I)$ due to phase fluctuations.  Together with recent work indicating giant $T_c$ enhancements in superconducting nanoparticles~\cite{Shanenko2006a,Tian2015,Kresin2006,Bose2010,He2013}, this inbuilt resilience to phase fluctuations supports the assembly of dirty nanowire arrays as an attractive route towards synthesizing new functional superconductors. Moreover, {\Na} and similar q1D filamentary materials provide unrivalled opportunities for investigating dimensional crossover and its influence on emergent electronic order: a field of key relevance to low-dimensional materials and nanostructures.

\section{Methods}
\subsection{Crystal growth and characterization} 
\scriptsize{Na$_2$Mo$_6$Se$_6$ precursor powder was prepared using a solid-state ion exchange reaction technique~\cite{Potel1984}.  {\Na} single crystals of mass approximately 150~$\mu$g and diameter 100-200~$\mu$m were grown by heating this cold-pressed powder in a sealed Mo crucible at 1750~$^\circ$C for 3 hours.  A full structural (X-ray) characterization may be found in a .cif file attached to the Supporting Information.}
\newline

\subsection{Transport measurements} 
\begin{figure}
\centering
\includegraphics [width=0.5\linewidth] {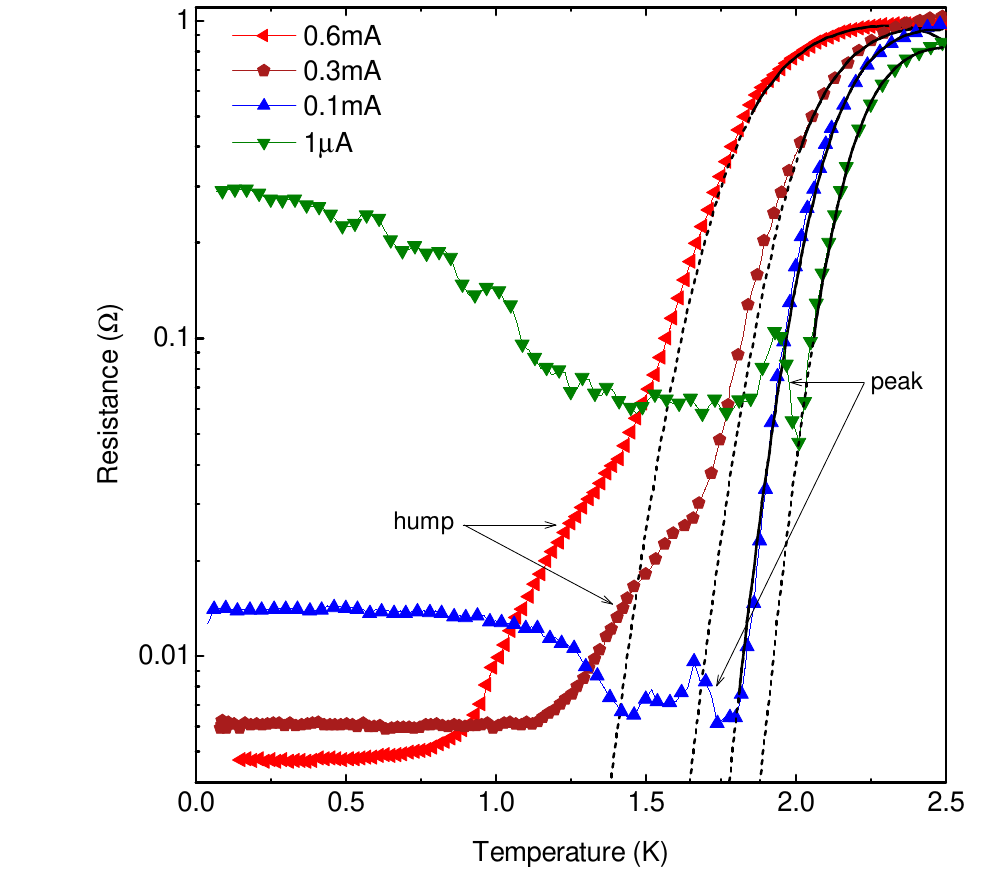}
\caption {\label{Fig2} $R(T)$ curves for $0.04$~K~$<T<2.5$~K and $I=1$~$\mu$A, $0.1$, $0.3$ and $0.6$~mA.  $2.5$~K is the maximum temperature attainable in our dilution refrigerator; $T_{ons}=2.7$~K was determined using $\left.dR/dT\right|_{T_{ons}}=0$ from a separate measurement in another cryostat (see Fig.~S5).  Data are fitted using a modified LAMH model over the range $T_{J}<T<T_{ons}$ (solid black lines, switching to dotted lines where the fits diverge from the experimental data).  The same fits are plotted on a linear $y$-scale in SI Fig.~S3(a) for comparison.  Only $50\%$ of our raw data-points are indicated in each $R(T)$ curve for clarity. 
}
\end{figure}

\begin{figure*}
\centering
\includegraphics [width=\linewidth] {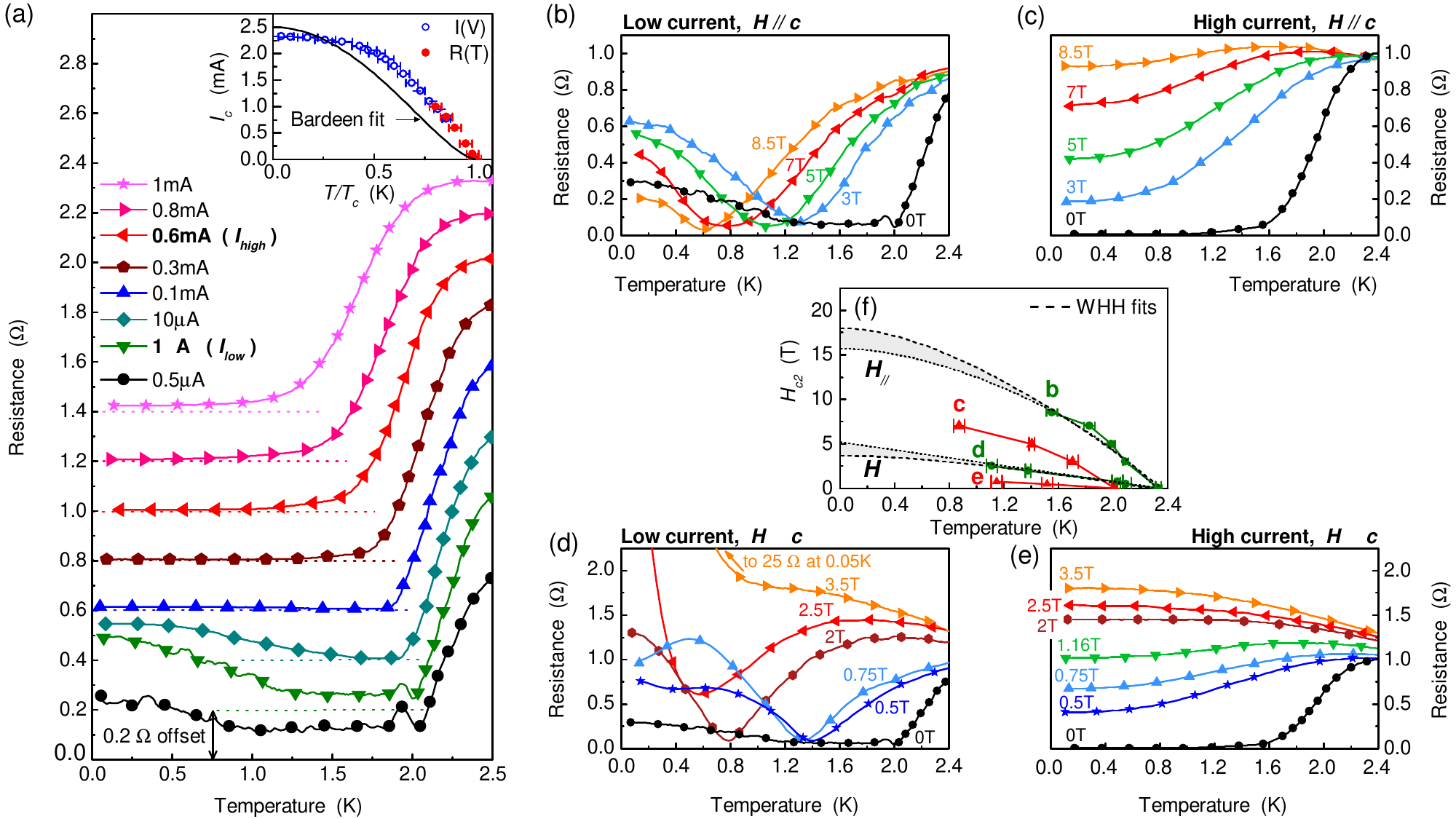} 
\caption {\label{Fig3} Reentrant phase coherence and the eventual suppression of superconductivity by current and magnetic field.  (a) $R(T)$ in zero magnetic field, acquired using 8 different currents $0.5\mu$A~${\leq}~I\leq1$~mA. An offset of $0.2\Omega$ separates each curve for clarity; dashed lines indicate $R=0$ for each data-set.  Inset: $I_c(T)$, defined as $I(R(T)=0.8R_{\mathrm{NS}})$, where $R_{\mathrm{NS}}$ is the normal-state resistance at $T_{ons}$.  Red circles correspond to data from Fig.~\ref{Fig3}(a), while blue data at larger $I_c$ are extracted from IV curves in Fig.~S3(a).  The solid line shows the theoretical $I_c(T)$ variation in a 3D material~\cite{Bardeen1962}.  (b-e) Evolution of $R(T)$ in magnetic fields applied parallel ($0{\leq}{H_{//}}{\leq}8.5$~T, (b,c)) and perpendicular ($0{\leq}{H_\perp}{\leq}3.5$~T, (d,e)) to the $c$ axis, using two excitations: $I_{\mathrm{low}}$~=~1~$\mu$A (b,d) and $I_{\mathrm{high}}=0.6$~mA (c,e).  (f) $H_{c2/\!/,\perp}(T)$ extracted from (b-e).  A similar $80\% R_{\mathrm{NS}}$ criterion is used to define $H_{c2}$, \textit{i.e.} $H_{c2}(T)~{\equiv}~H(R(T)=0.8R_{\mathrm{NS}})$.  We define $H_{c2}(T)$ (and $I_c(T)$) in this manner to consistently characterize the entire superconducting phase, since the error in $T_{ons}(H,I)$ is large for high $(H,I)$.  The dotted/dashed lines are fits representing upper/lower limits to $H_{c2}(T,I_\mathrm{low})$.  For $H_{c2\perp}$, the lower limit is defined by a WHH fit to the experimental data, while the upper limit assumes that $H_{c2}(T)$ remains linear at all temperatures, similar to a previous report for {\Tl}~\cite{Brusetti1994a}.  For $H_{c2/\!/}$, the limits are calculated using WHH fits to $H_{c2}(T\pm{\triangle}T)$, where ${\triangle}T$ is the error on the temperature axis.  
}
\end{figure*}

\begin{figure}
\centering
\includegraphics [width=0.5\linewidth] {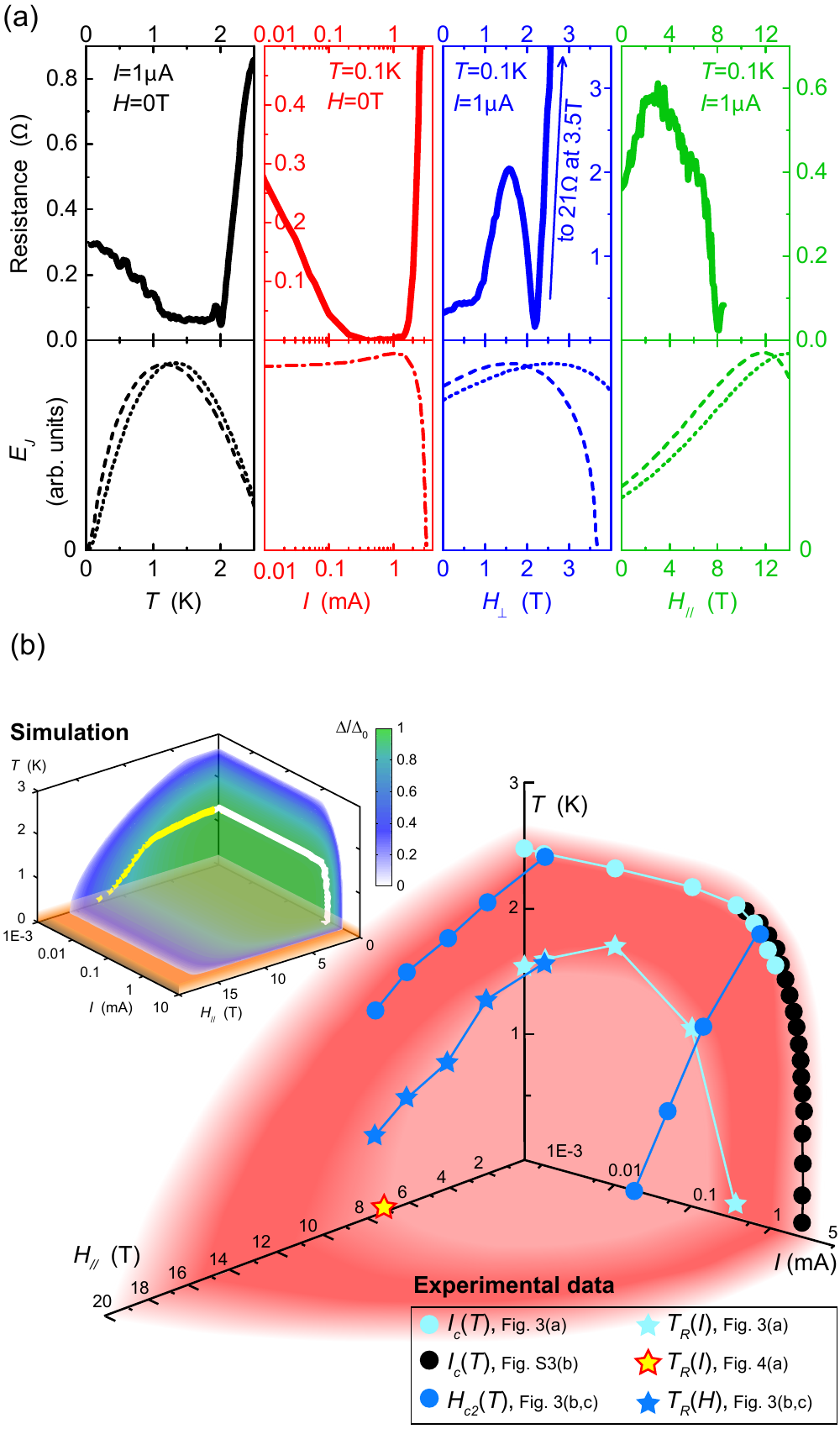} 
\caption{\label{Fig4} Evolution of reentrant coherence across ($T,H,I$) phase space.  (a) Resistance $R(T,I,H_\perp,H_{/\!/})$ in the superconducting phase (above, solid lines), together with our estimated coupling between superconducting filaments $E_J(T,I,H_\perp,H_{/\!/})$ (below).  Upper/lower limits in $E_J$ are denoted by dotted/dashed lines, respectively; the $E_J$ estimation is detailed in SI section VII.  When $E_J$ is maximized, phase coherence is established between filaments and hence the resistance is minimized.  (b) Experimental phase diagram illustrating $H_{c2},I_c$ (circles) and $T_R(H,I)$ (stars).  Dark red shading is a guide to the eye, highlighting the shell of reentrant phase coherence at elevated $(T,H,I)$, while the fluctuation-dominated incoherent region at low $(T,H_{/\!/},I)$ is shaded in pink.  Replacing $H_{/\!/}$ by $H_\perp$ leads to a similar diagram with the $H$ axis normalized by $\epsilon$.  Inset: calculated phase diagram showing $T_R(H,I)$ (yellow/white data-points, obtained from maxima in $E_J(T,H,I)$) and the normalized pairing energy $\triangle(T,H,I)/\triangle_0$ (blue-green shading).  The close correspondence between our calculated and experimental $T_R(H,I)$ illustrates how we may establish transverse phase coherence by increasing the strength of the Josephson coupling.  Despite providing realistic values for $T_R$, our basic model fails in the $T\rightarrow0$ limit since it predicts $E_J\rightarrow0$: the $T=0$ plane is shaded orange to highlight this deficiency.}
\end{figure}

\begin{figure}
\centering
\includegraphics [width=0.5\linewidth] {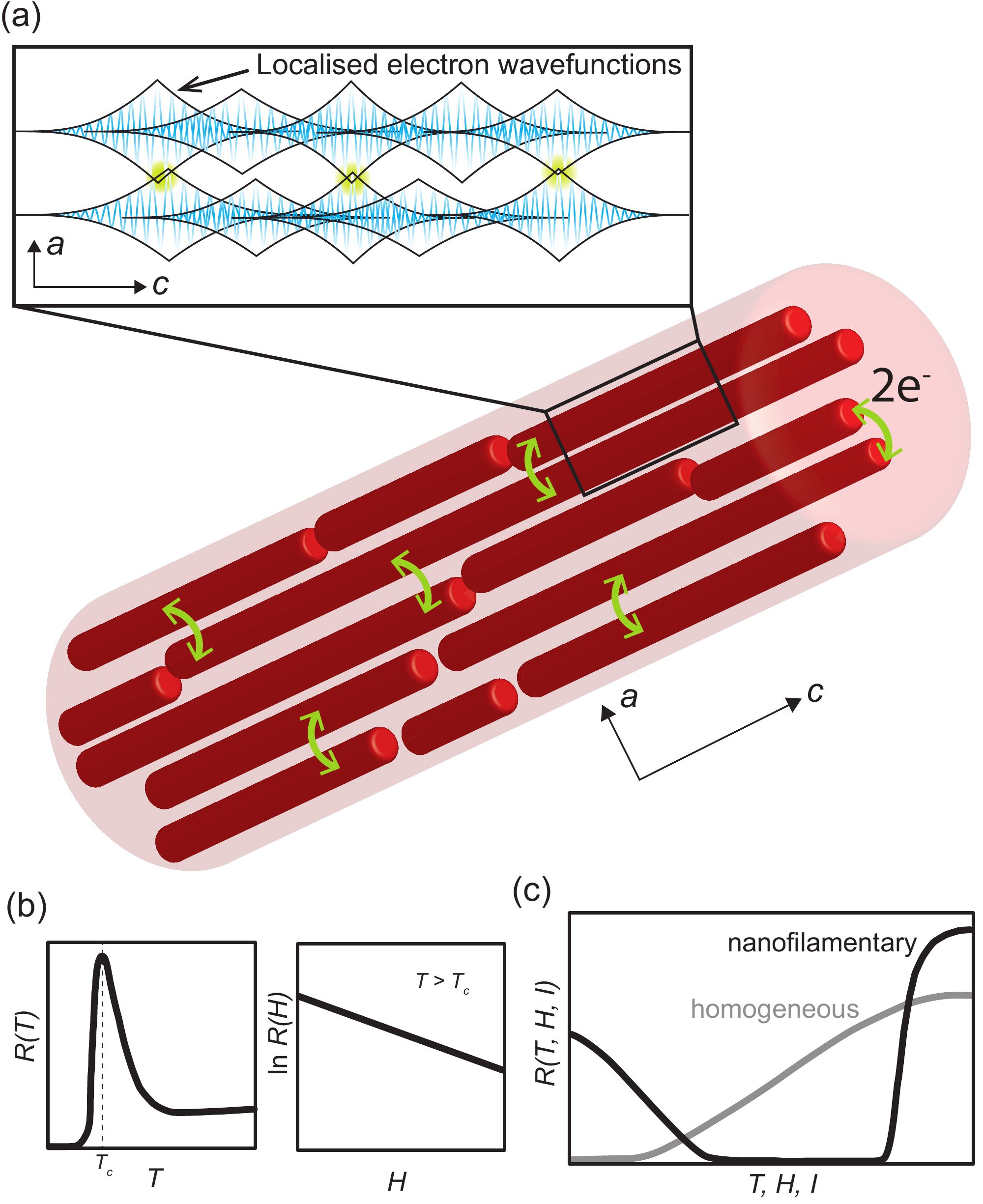} 
\caption{\label{Fig5} Inhomogeneous nanofilamentary materials as future functional superconductors.  (a) Schematic illustrating the nanoscopic structure of our proposed filamentary superconductor: in the normal state, the electron wavefunction overlap (green shading) between the filaments is small due to localization.  (b) Signatures of localization will be visible in the electrical transport: $dR/dT<0$ and $dR/dH<0$ for $T>T_c$.  (c) Electrons are delocalized as $(T,H,I)$ increase, enhancing the inter-filamentary coupling and leading to reentrant phase coherence: $R(T,H,I)$ therefore falls.  The result is a material whose superconducting properties improve for $(T,H,I)>0$, in contrast with conventional homogeneous superconductors (grey line) in which $R$ rises monotonically upon raising $(T,H,I)$.  Note that (b,c) are merely schematics illustrating typical normal-state transport properties required for reentrance and their effects on nanofilamentary superconductivity respectively.  For real $R_{\mathrm{NS}}(T,H)$ data from {\Na}, see Fig.~S5.}
\end{figure}

\scriptsize{Crystals were initially cleaned using sequential baths of hydrochloric acid, an ethanol/acetone mixture and distilled water.  Subsequently, four Au pads of thickness 20~nm were sputter-deposited onto the surface, two at each end of the crystal ($I_{+/-}$) and two closer to the centre ($V_{+/-}$).  Electrical contacts were made to these pads using 50~$\mu$m Au wire and Epotek E4110 Ag-loaded epoxy.  Resistivity measurements were performed using a standard ac four-probe method ($\nu=470$~Hz) using two separate systems: a Quantum Design Physical Property Measurement System (PPMS) with a 14~T magnet and a cryogen-free dilution refrigerator equipped with a 9~T/4~T vector magnet.  RF noise was removed from our measurement cables using ferrite filters prior to entering the dilution refrigerator.  Inside the refrigerator, all signals were carried by stainless steel microcoaxial cables.  To remove blackbody radiation, the cables were thermally anchored at numerous points (including the mixing chamber, \textit{i.e.} the coldest part of the refrigerator) before reaching the sample.  The standard inbuilt ac transport hardware was used in the PPMS, while $R(T)$ measurements in the dilution refrigerator were performed using a Keithley 6221 ac current source and a Stanford SRS830 lock-in amplifier.  Both methods provided identical and reproducible data.  Importantly, no phase shift was observed by the lock-in, implying that no extrinsic capacitive effects are present in our data.  The $V(I)$ curves in Fig.~\ref{Fig2} were acquired using a pulsed dc technique with a Keithley 6221 current source and 2182A nanovoltmeter.  The crystals are fragile and highly sensitive to thermal cycling from 0.05-300K.  They therefore exhibit a finite experimental lifetime, at the end of which $R_{NS}$ exhibits small irreversible jumps after each subsequent thermal cycle.  We exclude such ``end of lifetime data'' from our analysis.  The data which we plot in Figs.~2-4 are directly obtained from the raw voltage output of the lock-in using $R=V/I_{ac}$: no further data-processing is performed.}

\begin{acknowledgement}
We thank A. Chang and M. Croitoru for enlightening discussions, and Alexei Bosak (Beamline ID28, ESRF Grenoble) for assistance with data collection and processing.  This work was supported by the National Research Foundation, Singapore, through Grant NRF-CRP4-2008-04.\\

This document is the unedited Author's version of a Submitted Work that was subsequently accepted for publication in ACS Nano, copyright $\copyright$ American Chemical Society after peer review. To access the final edited and published work see DOI: 10.1021/acsnano.5b05450.
\end{acknowledgement}

\begin{suppinfo}

Supplementary data including electronic structure calculations, X-ray diffraction, modelling of q1D superconducting transitions, calculation of the filamentary diameter, crystal simulation using an anisotropic random resistor network, Werthamer-Helfand-Hohenberg fits for $H_{c2}$($T$), Josephson energy simulation, a discussion of additional reentrance mechanisms and the impact of tuning the disorder level. (PDF)\newline
Crystallographic Information File (CIF)\newline
\end{suppinfo}


\newpage

\clearpage



\section*{Supplementary material}

# SUPPORTING INFORMATION

## I. ELECTRONIC STRUCTURE OF $Na_2Mo_6Se_6$

We have performed *ab initio* density functional theory (DFT) calculations of the electronic structure of stoichiometric $Na_2Mo_6Se_6$, using the internal coordinates obtained from our X-ray structural refinement described in Sect. II and the full-potential linear augmented-plane-wave method.[1,2] The main features of the electronic structure are nearly identical to those previously calculated and published[3] for $Rb_2Mo_6Se_6$, $In_2Mo_6Se_6$, and $Tl_2Mo_6Se_6$. In Fig. S1 we show the band structures for $Na_2Mo_6Se_6$, $Rb_2Mo_6Se_6$ and $Tl_2Mo_6Se_6$, for **k** along the central $k_z$-axis (ΓA) and along lines (AL-LH-HA) perpendicular to it on the Brillouin-zone boundary ($k_z = \pi/c$). In a 1 eV region around the Fermi level, $E_F \equiv 0$, there is only *one*, spin-degenerate band, which crosses the BZ boundary at the Fermi level for the stoichiometric ($\delta = 0$) compounds. The band is *not* gapped at the BZ boundary because translation by $c/2$ followed by 180°-rotation around any $z$-axis is a covering operation of the crystal; choosing **k** to enumerate the irreducible representations of this Abelian group, rather than of the translation group, the band structure can be folded out to a BZ twice as high, whose boundary lies at $k_z = \pi/(c/2)$ and whose formula unit is $MMo_3Se_3$.[4]

The conduction band has a strong and linear dispersion in the 1 eV region around the Fermi level where there are no other bands. When *linearly* extrapolated and folded out to the distance $2\Gamma A = \pi/(c/2)$, the conduction band may be seen to have a width $W$ which is 7.4 eV for $Na_2Mo_6Se_6$, and nearly the same when $M = $ K, Rb, In and Tl (Table II in Ref.).[3] The corresponding velocity component is $v_\| = Wc/(2\pi)$ and, taking the dispersion to be $-2t_\| \cos(k_z c/2)$, the hopping integral is $t_\| = W/(2\pi) = 1.2$ eV for $Na_{2-\delta}Mo_6Se_6$ and nearly the same for the other cations (see Table I). This hopping along the chain is between neighboring $Mo_3$-molecular orbitals. Each of these is the antibonding linear combination of the 3 atomic $4d_{xz}$-orbitals on the $Mo_3$-triangle and $x$ is the local tangential direction.

In the direction perpendicular to the chain, the dispersion is very small. From Fig. S1 we see that for $Na_2Mo_6Se_6$, $Rb_2Mo_6Se_6$, and $Tl_2Mo_6Se_6$, the warping is $w = 90, 23$, and 180 meV respectively. Since the relative energies at A, L, and H are respectively $-6t_\perp, 2t_\perp$, and $3t_\perp$ in terms of the hopping integral between neighboring chains, $t_\perp = w/9 = 10, 3,$ and 20 meV for $Na_2Mo_6Se_6$, $Rb_2Mo_6Se_6$, and $Tl_2Mo_6Se_6$. This strong material dependence comes from the facts that the inter-chain distance is larger for the Rb than for the Na compound and that the hopping proceeds *via* the $M$-cation, and less so through polarization of $M$-valence orbitals with $s$ than with $p$-character (Fig. S1).

Note that $w/W$ is not quite as tiny as $t_\perp/t_\|$ since in the perpendicular direction there are 6 nearest-neighbor chains, but in the parallel direction only 2 nearest-neighbor triangles. The tight-binding expression for the band dispersion is[5]:

$$E(\mathbf{k}) = -2t_\| \cos\left(\frac{c}{2}k_z\right) - 2t_\perp \left\{\cos(ak_y) + \cos\left(\frac{a}{2}\left(k_y + \sqrt{3}k_x\right)\right) + \cos\left(\frac{a}{2}\left(k_y - \sqrt{3}k_x\right)\right)\right\};$$

Here, in contrast to the convention used to define the molecular orbitals above, the cartesian $xyz$-system is global with $y$ pointing between nearest-neighbor chains.

With $w/W$ being so small, the density of states (DoS) is simply $N(E) = 1/W$ states per spin per $MMo_3Se_3$, independently of $w$, $E$, and the $M$-stoichiometry, as long as the dispersion is linear and no other band conducts. In the rigid-band band picture, this is valid from approximately $M_{0.75}$ to $M_{1.05}$. The measured Na content of our crystals ($\delta \sim 0.2$) lies comfortably within this range. For $Na_{2-\delta}Mo_6Se_6$ and $Tl_2Mo_6Se_6$, respectively, we therefore obtain $N(E) = 0.135$ and 0.166 states/(eV×spin×$MMo_3Se_3$). This bare DoS will be renormalized by electron-phonon and electron-electron interactions with a mass-enhancement factor $1+\lambda$. This can be obtained from the ratio between the experimental electronic specific heat coefficient $\gamma$ and the bare DoS. In the absence of heat capacity data for $Na_{2-\delta}Mo_6Se_6$, we use data[3] from $In_2Mo_6Se_6$ and, assuming that the mass enhancement is the same for the two materials, we end up with an enhanced DoS of 0.17 states/(eV×spin×$MMo_3Se_3$) for $Na_{2-\delta}Mo_6Se_6$.

For the present study, the most important parameters from the electronic structure are $t_\|$ and $t_\perp$. Our calculated values (in °K) for the Na, K, Rb, In, and Tl chain compounds are given in Table I. For comparison, we also provide the hopping integrals for two of the best-known q1D superconductors: the Bechgaard salt $(TMTSF)_2ClO_4$ and the purple bronze $Li_{0.9}Mo_6O_{17}$, as well as some recent experimental data from the highly one-dimensional $Sc_3CoC_4$.

These calculations provide a significant advance in understanding the superconducting transition in q1D materials. Specifically, the relative size of $t_\perp$ and $t_{/\!/}$ deter-

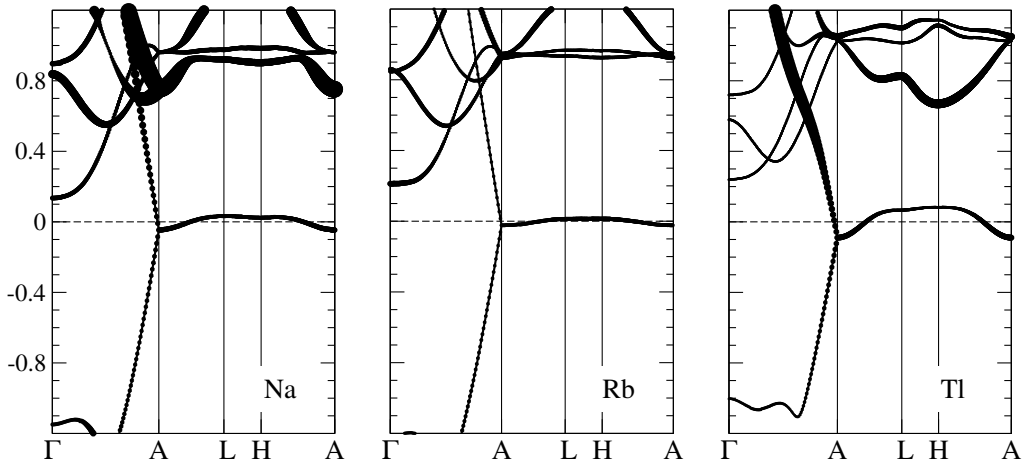

Figure S 1: Calculated band structures of $M_2Mo_6Se_6$ for $M$ = Na, Rb, and Tl decorated with $M$ valence-orbital characters. Energies are in eV with respect to the Fermi level of the stoichiometric compounds.

TABLE I: **Electronic anisotropy and superconducting transition temperatures in q1D materials**
Hopping integral units have been converted from eV to K for ease of comparison. $In_2Mo_6Se_6$,[3] $(TMTSF)_2ClO_4$[6] and $Li_{0.9}Mo_6O_{17}$[7] all exhibit single, sharp and complete superconducting transitions at $T_c$.

|  | $t_{//}$ | $t_\perp$ | $t_\perp^2/t_{//}$ | $T_c$ | $T_{ons}$ | $T_{BKT}$ | $t_\perp^2/t_{//}T_P$ |
|---|---|---|---|---|---|---|---|
| $Tl_2Mo_6Se_6$ | 12000 | 230 | 4.4 | – | 6.5 | 4.5 | 0.7 |
| $In_2Mo_6Se_6$ | 12000 | 190 | 3.0 | 2.85 | – | – | 1.05 |
| $Na_{2-\delta}Mo_6Se_6$ | 14000 | 120 | 1.0 | – | 2.7 | 1.73 | 0.4 |
| $K_2Mo_6Se_6$ | 14000 | 60 | 0.3 | – | – | – | – |
| $Rb_2Mo_6Se_6$ | 14000 | 30 | 0.06 | – | – | – | – |
| $(TMTSF)_2ClO_4$[8] | 3000 | 300 | 30 | 1.4 | – | – | 21 |
| $Li_{0.9}Mo_6O_{17}$[9] | 9300 | 170 | 3.3 | 1.9 | – | – | 1.7 |
| $Li_{0.9}Mo_6O_{17}$[10] | 8600 | 230 | 6.2 | 1.9 | – | – | 3.3 |
| $Sc_3CoC_4$ |  |  |  |  | 4 - 4.5 | 1.55 |  |

mines whether a single, sharp transition to the superconducting state occurs, or if the material exhibits a 2-step transition. In the case of a 2-step transition, a fluctuating phase-incoherent 1D superconducting phase is initially formed at $T_J < T < T_{ons}$ and phase coherence is only established below $T_J$. So far, 2-step transitions have been reported only in $Na_{2-\delta}Mo_6Se_6$, $Tl_2Mo_6Se_6$[11] and $Sc_3CoC_4$.[12]

If we compare data from the literature in other q1D superconductors, we find that a crossover from single to 2-step transitions occurs at $t_\perp^2/t_{//}T_P = 1$. Here $T_P$ is the pairing temperature: in the absence of any pseudogap or other exotic effects, $T_P \equiv T_c$ for a conventional superconductor, while $T_P \equiv T_{ons}$ for a q1D material exhibiting a 2-step transition. This ratio is tabulated in the last column of Table I: we clearly see that a 2-step transition occurs for $t_\perp^2/t_{//}T_P < 1$.

A simple physical explanation exists for this crossover: $t_\perp^2/t_{//} \sim T_J$, where $T_J$ is the two-particle hopping – i.e. Josephson coupling – temperature. At temperatures above $T_J$, Josephson tunnelling of Cooper pairs between individual 1D filaments cannot occur and hence it is impossible to establish inter-filamentary phase coherence. Therefore, if $T_P > T_J$, a 2-step transition occurs and phase coherence is only established at $T_J$.

To the best of our knowledge, $M_2Mo_6Se_6$ and $Sc_3CoC_4$ are unique amongst q1D crystalline superconductors in possessing a sufficiently large electronic anisotropy to exhibit 2-step transitions. No calculations for the hopping integrals in $Sc_3CoC_4$ exist in the literature and we were therefore unable to verify whether this material complies with our empirical criterion for the crossover.

The key point which we wish to reinforce is that from a superconducting perspective, $Na_{2-\delta}Mo_6Se_6$, $Tl_2Mo_6Se_6$ and $Sc_3CoC_4$ all remain one-dimensional until $T < T_J$, upon which pairs of electrons may hop between filaments. The unique property which differentiates $Na_{2-\delta}Mo_6Se_6$ from the other compounds is its propensity for intrinsic disorder, caused by the large Na vacancy population. It is this disorder which leads to the reentrant behavior in the inter-filamentary phase coherence which we observe.

## II. SINGLE CRYSTAL X-RAY DIFFRACTION

X-ray diffraction experiments were performed at the Swiss-Norwegian Beamlines (SNBL) of the European Synchrotron Radiation Facility (Grenoble, France) at the end station BM01A, using a PILATUS2M pixel area detector.[13] A monochromatic beam at a wavelength of 0.694Å was slit-collimated down to a size of 100×100 $\mu m^2$. The sample-to-detector distance and the parameters of the detector were calibrated using a $LaB_6$ NIST standard. The detector images were recorded by $\phi$-scans in shutter-free mode with a 0.1° angular step. We have performed measurements both at room temperature (293 K) and 20 K, cooling the crystals using a helium blower. Our data were preprocessed by the SNBL Tool Box,[14] followed by the CrysAlis Pro[15] software package. The crystal structure was solved with SHELXS and subsequently refined with SHELX.[16] A crystallographic information (.cif) file acquired from a typical sample at room temperature is attached to this Supporting Information.

Our refinement indicates a hexagonal lattice with space group $P6_3/m$ and parameters $a = 8.65$Å, $c = 4.49$Å at 293 K, with a minimum inter-chain Mo-Mo separation of 6.4 Å. This is in agreement with previous reports.[17,18] The lattice parameters fall to $a = 8.61$Å, $c = 4.48$Å at 20 K. There is no evidence for any Peierls-type structural transition (i.e. a $2k_F$ charge density wave doubling the c-axis lattice parameter) in the XRD patterns acquired at 20 K. This implies that the upturn in the resistance at low temperature (Fig. S6) is predominantly a consequence of electron localization.

The key result from our XRD experiments is the large Na deficiency observed in superconducting $Na_{2-\delta}Mo_6Se_6$: we measure $\delta = 0.2 \pm 0.036$ for the crystal whose .cif file is attached, while similar crystals provided $\delta = 0.22 \pm 0.030$ and $\delta = 0.26 \pm 0.08$. The quality of the refinement can be quantified using a goodness of fit parameter $\chi_f$ provided by SHELX, where $\chi_f = 1$ describes a flawless correspondence to the theoretical structure. We obtain $\chi_f = 1.086$, 1.160 and 1.351 respectively for the three crystals detailed above. It is therefore clear that although small variations may exist between crystals, the typical Na deficit is at least 10%. Crucially, we cannot detect any Mo or Se deficit, implying that the $(Mo_6Se_6)_\infty$ chains remain highly structurally ordered.

## III. MODELLING Q1D SUPERCONDUCTING TRANSITIONS

### A. Estimating $\xi(0)$

Before proceeding with any quantitative analysis of the superconducting transition in $R(T)$, we require an estimate of the coherence length $\xi$. We can of course use our experimentally-determined $\xi_{//}$ from our fits to the upper critical fields $H_{c2//,\perp}$ in Fig. 3(f) from the main text, but it is instructive to compare this value with that obtained from dirty-limit BCS theory, $\xi_d$.

A superconductor lies in the dirty limit if $l < \xi_0$, where $l$ is the mean free path and $\xi_0$ is the clean limit coherence length. The substantial presence of Na vacancies suggests that $Na_{2-\delta}Mo_6Se_6$ will fall into this category: to verify this, we evaluate $l$ using $1/\rho = ne^2 l/\hbar k_F$ (where $n = 3.13 \times 10^{21}$ cm$^{-3}$ is the carrier density assuming $0.9 e^-$/unit cell, i.e. $\delta = 0.2$) and $\rho$ is the normal-state resistivity. To avoid any extraneous influence from localization, we use $\rho(300\ K) = 1.1 \times 10^{-6}$ $\Omega$m, yielding $l \sim 8$ nm. $\xi_0$ is calculated from the BCS relation $\xi_0 = \hbar v_F / 1.76 \pi k_B T_c$, in which we set $T_c \equiv T_{ons}$ and $\hbar v_F = dE/dk|_{k_F} \equiv 3.6$ eV × $c/\pi$ (from our band structure calculations). This yields $\xi_0 \lesssim 397$ nm $\gg l$. The clean limit $\xi_0$ must therefore be replaced by the dirty limit $\xi_d = 0.85\sqrt{l\xi_0}$ and we hence obtain $\xi_d = 48.0$ nm. Our experimentally-estimated $\xi_{//} = 14.4$-21.0 nm lies below this BCS value, as expected for a short coherence length superconductor.[19]

### B. Introducing the LAMH Model

In a 1D filament at temperature $T < T_{ons}$ (where $T_{ons}$ is the pairing temperature and hence corresponds to the onset of superconducting fluctuations), the superconducting order parameter may fluctuate to zero at certain points along the filament. This allows the phase to slip by $2\pi$, creating a resistive state. A theory describing these thermally activated phase slips was originally developed by Langer, Ambegaokar, McCumber and Halperin (LAMH)[20,21]: within this model, phase slip formation necessitates overcoming an energy barrier $\triangle F$ proportional to the superconducting condensation energy, the coherence length $\xi(T) = \xi(0)(1 - T/T_{ons})^{-1/2}$ and the cross-sectional area of the wire. A characteristic timescale for the fluctuations is fixed using a prefactor $\Omega$, related to the attempt frequency of random excursions in the superconducting order parameter. The resultant fluctuation-dominated resistance of a 1D superconducting wire can be expressed as follows:

$$R_{LAMH}(T) = \frac{\pi \hbar^2 \Omega}{2e^2 k_B T} \exp\left(\frac{-\triangle F}{k_B T}\right) \quad (1)$$

where the attempt frequency is given by:

$$\Omega = \frac{L}{\xi}\left(\frac{\triangle F}{k_B T}\right)^{1/2} \frac{1}{\tau_{GL}} \quad (2)$$

and $\tau_{GL} = [\pi \hbar / 8 k_B (T_{ons} - T)]$ is the Ginzburg-Landau relaxation time. Following a development of the energy barrier by Lau et al.,[22] we write $\triangle F$ as

$$\triangle F = C k_B T_{ons}\left(1 - \frac{T}{T_{ons}}\right)^{3/2} \quad (3)$$

where $C$ is a dimensionless parameter relating the energy barrier for phase slips $F$ to the thermal energy near $T_{ons}$:

$$C \approx 0.83 \left(\frac{L}{\xi(0)}\right)\left(\frac{R_q}{R_{NS}}\right) \quad (4)$$

Here $R_q = h/4e^2 = 6.45$ $k\Omega$ is the resistance quantum for Cooper pairs and $R_{NS}$ the normal state resistance of the wire. We plot the temperature dependence of $\triangle F$ and $\Omega$ in Figs. S2(a,b): both parameters decrease to zero as $T \to T_{ons}$. However, $\triangle F$ behaves as an activation energy and $R_{LAMH}$ therefore falls exponentially as $T \to 0$, reaching zero for $T \lesssim 0.5 T_{ons}$ (Fig. S2(c)). Finally, the total resistance of the 1D filament is evaluated by including the normal-state quasiparticle contribution, which is assumed to be temperature-independent for simplicity:

$$R = (R_{NS}^{-1} + R_{LAMH}^{-1})^{-1} \quad (5)$$

It is important to note that the peak emerging in $R_{LAMH}$ and $R$ (Figs. S2(c,d)) is a mathematical artifact with no physical significance, which should be disregarded during the fitting procedure.[23–25]

### C. Applying LAMH theory to $Na_{2-\delta}Mo_6Se_6$

A macroscopic $Na_{2-\delta}Mo_6Se_6$ crystal cannot be treated as a single quantum wire within LAMH theory since its diameter is much greater than $\xi$. Instead, we model a $Na_{2-\delta}Mo_6Se_6$ crystal as a $m \times n$ array of 1D filaments, i.e. a $m \times n$ network of identical resistors $R_F$ (Fig. S4(a)). The choice of a 2D array to describe a crystal existing in 3 spatial dimensions is made purely for simplicity and has no effect on our results, since the resistance of a line of $n$ resistors in parallel is equivalent to that of a $\sqrt{n} \times \sqrt{n}$ lattice. The total normal-state resistance of our $m \times n$ array is $R_{NS} = mR_F/n$. Below $T_{ons}$, $R^{-1} = R_{NS}^{-1} + \frac{n}{m} R'^{-1}$ where $R'$ is the LAMH resistance of a single superconducting filament of length $L$, where $C \approx 0.83 \left(\frac{L}{\xi(0)}\right)\left(\frac{R_q}{R_F}\right)$. Now, we would like to re-express $C$ in terms of $R_{NS}$ (which we know) and therefore write $C \approx 0.83 \left(\frac{Lm}{n\xi(0)}\right)\left(\frac{R_q}{R_{NS}}\right)$. Defining an effective length $L_{eff} = Lm/n$ and renormalizing $\Omega$ accordingly, we obtain $R^{-1} = R_{NS}^{-1} + R_{eff}^{-1}$. $R_{eff}$ is the total LAMH contribution to the resistance, controlled by $C \approx 0.83 \left(\frac{L_{eff}}{\xi(0)}\right)\left(\frac{R_q}{R_{NS}}\right)$.

The above argument shows that standard LAMH theory using a geometrically renormalized length $L_{eff}$ can describe q1D superconducting crystals as well as individual nanowires, with the limiting assumption that all the filaments in the crystal have the same geometry. We therefore perform least-squares fits to our experimental $R(T)$ curves using equation 5, with $L_{eff}/\xi(0)$ and $T_{ons}$ as free parameters. The results are summarized in Table II, while the fits are displayed graphically in Fig. 2(a) from the main text. We note that the rise in $\xi(0)$ with increasing current which we infer from these fits is also clearly revealed by the reduction in $H_{c2}$ at high currents (Fig. 3(f) in the main text).

### D. Indicators for the onset of transverse phase coherence below the LAMH regime

At low currents $I \lesssim 0.1$ mA, a sharp peak in $R(T)$ emerges at the lower boundary of the temperature range

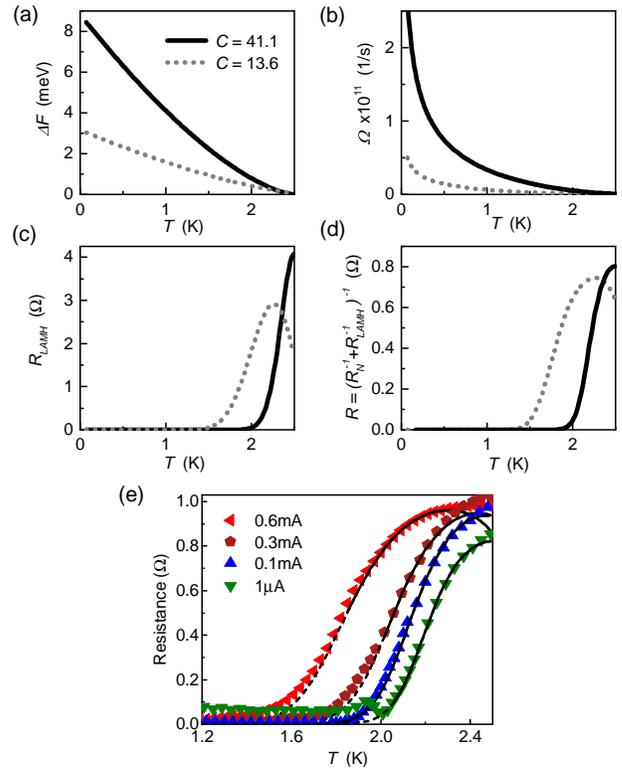

Figure S 2: (a-d) Parameter evolution with temperature in the LAMH model for 1D superconducting transitions. All curves are calculated using $R_{NS} = 1$ $\Omega$ and $T_{ons} = 2.7$ K, similar to the values observed in our experiments. Each parameter is evaluated using $C = 41.1$ (black solid line) and $C = 13.6$ (grey dotted line), to illustrate the influence of the $L/\xi(0)$ ratio. (a) $\triangle F$ from equation 3; (b) $\Omega$ from equation 2; (c) $R_{LAMH}$ from equation 1; (d) Total sample resistance $R = (R_{NS}^{-1} + R_{LAMH}^{-1})^{-1}$; (e) LAMH fits (black lines) to experimental $R(T)$ data acquired at four distinct excitation currents, plotted on a linear y-scale (which masks the "hump" developing close to $T_J$). The fits to $I = 0.6$ mA and $I = 1$ $\mu$A correspond to the values $C = 41.1$ and $C = 13.6$, respectively. Data are identical to those shown in Fig. 2 of the main text.

within which the LAMH model is valid (Fig. 2 of the main text). We interpret this feature as follows: upon reducing the temperature, the superconducting gap is growing and the quasiparticle density of states falls, thus suppressing quasiparticle transport between filaments. Eventually, this quasiparticle reduction balances the drop in $R(T)$ due to fluctuating 1D superconductivity within individual filaments, causing a minimum and subsequent upturn in $R(T)$. The upturn is abruptly suppressed by the onset of transverse phase coherence at $T_J$, which enables Cooper pair transfer between $(Mo_6Se_6)_\infty$ filaments and hence establishes long-range superconducting order: $R(T)$ therefore falls once more, creating a peak. As the applied current rises, we anticipate that pair-breaking should increase the quasiparticle population close to $T_J$, thus reducing the size of the upturn in $R(T)$ and gradu-

TABLE II: **LAMH fitting parameters and their evolution with current in a $Na_{2-\delta}Mo_6Se_6$ single crystal.** The errors quoted for $L_{eff}/\xi(0)$ and $T_{ons}$ are the standard deviations obtained from our least-squares fitting routine. Fitting ranges define the segments of each $R(T)$ curve used to perform each fit.

| $I$ | $L_{eff}/\xi(0)$ | $T_{ons}$ (K) | Fitting range (K) |
|---|---|---|---|
| 1 µA | 0.0082±0.0004 | 2.73±0.02 | 2-2.45 |
| 0.1 mA | 0.0075±0.0004 | 2.74±0.02 | 1.8-2.4 |
| 0.3 mA | 0.0057±0.0002 | 2.76±0.02 | 1.71-2.4 |
| 0.6 mA | 0.0037±0.0002 | 2.71±0.02 | 1.71-2.3 |

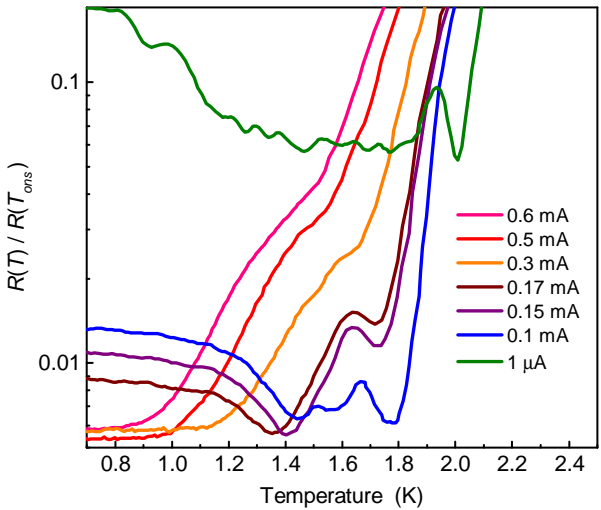

Figure S 3: $R(T)$ data normalized to the normal-state resistance at $T_{ons}$ for currents varying from 1 µA to 0.6 mA. As $I$ increases, the finite-resistance plateau vanishes and the peak terminating the 1D LAMH regime at higher temperature is gradually broadened into a hump.

ally smearing the peak into a broad hump.

Figure S3 shows an extended series of $R(T)$ curves illustrating the progression from peak to hump with increasing $I$. In particular, we highlight the data acquired at $I = 0.15, 0.17$ mA, where $R(T)$ exhibits a small peak at $\sim 1.65$ K. Below this peak $R(T)$ continues to fall, but eventually passes through a minimum at $T_R$ and rises again thereafter. The "reentrance threshold" $T_R(H,I)$ falls with increasing current, disappearing for $I \geq 0.3$ mA. These data support our assertion that the peak/hump and low-temperature rise in $R(T)$ correspond to the onset and loss of transverse phase coherence, respectively. Furthermore, the presence of these two features is clearly independent from (and hence unrelated to) the finite-resistance plateau visible for $I < 10$ µA. We believe that this plateau is masking the presence of long-range superconducting order at low current: its likely origins will be discussed further in section VIII.

### E. Similarities to the 2D BKT transition

Although Berezinskii-Kosterlitz-Thouless (BKT) transitions are generally considered to be a property of 2D materials, evidence for BKT-type behavior has been reported in experimental and numerical studies of q1D superconductors.[11,12,30,31] In Fig. S4(a), we sketch the spatial variation of the phase of the superconducting order parameter on each $(Mo_6Se_6)_\infty$ filament in a cross-sectional "slice" perpendicular to the $c$ (high-symmetry) axis. For $T > T_J$ in a q1D superconductor, phase fluctuations within individual filaments will take place over distances of the order of $\xi_{//}$. In $Na_{2-\delta}Mo_6Se_6$, $\xi_{//}$ is almost two orders of magnitude larger than the characteristic lengthscale for transverse phase fluctuations (the inter-filamentary separation). This large anisotropy suggests that the influence of phase fluctuations along the $c$ axis may be weak, thus encouraging us to visualize q1D crystals as arrays containing many such slices of thickness $\xi_{//}$ or more, in which phase fluctuations parallel to $c$ may be ignored. A clear analogy between q1D and 2D materials now emerges: within an individual slice, the phase of the order parameter on each filament satisfies 2D XY symmetry. In this scenario, the BKT transition temperature $T_{BKT}$ would be equivalent to $T_J$, at which phase-locking occurs between the filaments. We therefore examine our data for possible signatures of a BKT-type transition.

Several unique electrical transport characteristics are observable at BKT transitions. Firstly, $V(I)$ should exhibit power-law behavior with $V \propto I^{\alpha(T)}$, where $\alpha(T)$ is related to the superfluid density $\rho_s$ by $\alpha = 1 + \pi\rho_s\hbar^2/4m_ek_BT$. $\rho_s$ (and hence $\alpha$) rises steeply in a "Nelson-Kosterlitz jump" at the BKT transition,[26,27] with $\alpha(T_{BKT}) = 3$. Secondly, the resistivity scales exponentially as $R(T) = R_0 \exp(-bt^{-1/2})$ over a narrow temperature range above $T_{BKT}$, where $t = T/T_{BKT} - 1$ and $R_0, b$ are material constants.[28,29,32] Close to $T_{BKT}$, finite size effects limit the exponential divergence of the correlation length, creating a hump in $\log(R(T))$, which is further broadened below $T_{BKT}$ by the effects of inhomogeneity and vortex unbinding at elevated current.[33] [Such factors are likely to share responsibility with pair-breaking effects for the peak→hump smearing shown in Fig. S3.]

Together, these characteristics enable the accurate determination of $T_{BKT}$ from transport experiments, as previously demonstrated in a range of 2D[27,28,32,34] and suggested in q1D[11,12,30] materials. Figure S4(b) shows the evolution of $V(I)$ with temperature in $Na_{2-\delta}Mo_6Se_6$: a supercurrent gap opens for $T \lesssim 1.8$ K. The $V(I)$ curves display power-law behavior in the 1-2.5 mA range (Fig. S4(c)) with an exponent $\alpha(T) = 3$ at $T_{BKT} = 1.72 \pm 0.01$ K (Fig. S4(d)). $R(T)$ also displays a narrow exponential scaling regime from $\sim 1.8$-1.9 K (Fig. S4(e)) which may be extrapolated to obtain $T_{BKT} = 1.69 \pm 0.02$ K, in close agreement with the $\alpha = 3$ definition. We therefore conclude $T_{BKT} \equiv T_J = 1.71 \pm 0.02$ K. As discussed in the main text, we attribute the enhancement of $T_J$ relative to $t_\perp^2/t_{//} = 1$ K to the large Na vacancy

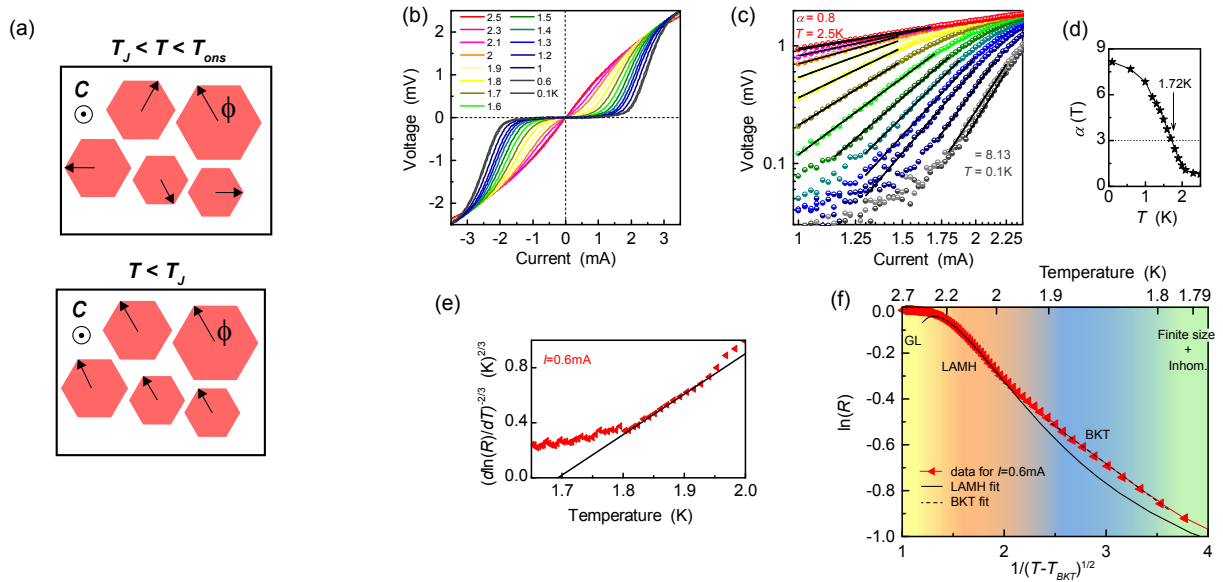

Figure S 4: (a) Emergence of 2D $XY$-symmetry in a q1D system within the plane perpendicular to the chains (c-axis). The correspondence with 2D BKT physics is apparent from the phase $\phi$ of the order parameter on each filament: above $T_J$ neighboring filaments are phase incoherent and fluctuating 1D superconductivity develops on individual filaments. Below $T_J$, phase-locking occurs between the filaments inducing a 1D→q3D dimensional crossover. (b) $V(I)$ curves for $T = 0.1$ to $2.5$ K, acquired using 90 $\mu$s dc current pulses. At high currents for $T \gtrsim 2.4$ K, non-Ohmic behavior ($\alpha < 1$) is due to sample heating, since $dR/dT < 0$ in the normal state (Fig. S5). (c) Power-law fitting to the $V(I)$ raw data from (b), where $V \propto I^{\alpha(T)}$ (solid lines). Within a BKT-type scenario, the Ohmic "tails" at low current can be created by finite size effects. (d) $\alpha(T)$ obtained from the fitting procedure in (c), displaying a Nelson-Kosterlitz jump. The sharp jump predicted[26] in $\alpha(T)$ at $T_{BKT}$ is substantially broadened by inhomogeneity in real materials.[27,28] (e) Exponential $R(T)$ scaling above $T_J$: extrapolating the linear region to zero (black line) yields $T_{BKT} = 1.69 \pm 0.02$ K. (f) Halperin-Nelson[29] rescaled resistance $\ln R$ vs. $1/(T - T_{BKT})^{1/2}$ for $I = 0.6$ mA, illustrating the principal contributions to $R(T)$ for $T_J < T < T_{ons}$. The solid black line is the LAMH fit from (a), while the dashed black line indicates BKT-type exponential scaling. Upon reducing the temperature, the LAMH fit starts to deviate from our data as we enter the narrow BKT-type regime.

disorder and possible strong correlation effects. For comparison, $Tl_2Mo_6Se_6$ crystals consistently exhibit a lower (2.5-5%) Tl deficiency[35] and display a closer correspondence between $T_J \equiv T_{BKT} = 4.5$ K[11] and $t_\perp^2/t_\parallel = 4.4$ K.

We summarize the sequential contributions to $R(T < T_{ons})$ in Fig. S3(f): close to $T_{ons}$, the usual Ginzburg-Landau (GL) fluctuations of the superconducting order parameter are replaced by LAMH phase slips as the temperature falls. LAMH theory is only valid in the 1D limit, i.e. at temperatures above the dimensional crossover at $T_J$. Approaching $T_J$ from above, the LAMH behavior gives way to a narrow regime of exponential BKT-type scaling close to $T_J$.

Given that exponential BKT-type scaling in $R(T)$ and LAMH phase slips may both contribute to $R(T)$ above $T_J$, one might wonder why the exponential component is not obscured by phase slips. In Fig. S3(f), we also include a Halperin-Nelson logarithmic rescaling[29] of the same LAMH $R(T)$ fit used to model our data acquired at $I = 0.6$ mA (Fig. 2(a)). A quasi-linear trend is visible over a broad temperature range, thus explaining why the linear BKT-type regime emerging at the lower end of this range can remain experimentally accessible. The divergence between the LAMH fit and our experimental data also becomes clear: while the LAMH fit rescales to a gentle curve in the $2.5 \lesssim 1/(T - T_J)^{1/2} \lesssim 3.5$ region corresponding to $1.9 > T > 1.8$ K, a linear BKT-type regime is observed in our rescaled data (Figs. S3(e,f)). BKT-type scaling is only visible in a narrow temperature range close to $T_J$, limited from above by phase slips and from below by finite size effects and inhomogeneity.

The quasi-linear behavior exhibited by the LAMH model after Halperin-Nelson rescaling is a consequence of the duality between vortices in a BKT transition and phase slips in 1D. In fact, a chain of Josephson junctions (analogous to a 1D array of phase slips) may also undergo a zero-temperature transition of the BKT universality class.[36,37] However, any BKT-type transition in $Na_{2-\delta}Mo_6Se_6$ would correspond to a 1D→q3D dimensional crossover rather than 1D phase slip binding, for numerous reasons beyond the fact that our experiments are performed at finite temperature. Although the 1D nature of superconductivity for $T_J < T < T_{ons}$ is indisputable since the LAMH model accurately describes $R(T)$, $Na_{2-\delta}Mo_6Se_6$ is a macroscopic crystalline material and must undergo a finite-temperature dimensional

crossover (at a minimum temperature of $t_\perp^2/t_{//} \sim 1$ K, if we disregard the effects of localization). The presence of 3D coupling at low temperature is also evident from the field-induced Cooper pair localization, which decouples the filaments and confines electron pairs on individual $(Mo_6Se_6)_\infty$ chains (Fig. 3(d) in the main text). Furthermore, in the absence of 3D ordering one would expect a quantum phase slip contribution[23–25] and an exponential decay in $R(T)$, leading to a finite resistance as $T \to 0$: such features are absent from our data.

## IV. CALCULATING THE FILAMENTARY DIAMETER

The results of our LAMH fitting procedure allow us to estimate the typical diameter $d_F$ of the superconducting filaments in $Na_{2-\delta}Mo_6Se_6$. First, let us naively assume that current flows homogeneously through the crystal. By setting $Lm = w$ (where $w = 0.4$ mm is the voltage contact separation) and using our experimental $\xi_{//} = 14$–21 nm, we may deduce $n$ (the number of filaments in a typical crystal cross-section) and hence estimate the filament diameter. For $I = 1$ μA, we obtain $n = 3.5$–$2.3 \times 10^6$ and use the crystal cross-section $A \sim 8 \times 10^{-9}$ m$^2$ to deduce a maximum filament diameter $d_F = 54$–66 nm. The diameter of a 1D superconducting filament must be smaller than its coherence length and so this figure is clearly far too large: we have overestimated $d_F$ due to the inhomogeneous current flow through the crystal. In such a strongly 1D material, defects will force charge transport to occur *via* a highly percolative route. The typical current-carrying fraction of the crystal cross-section will therefore be small ($\ll A$), leading to a considerable reduction in $d_F$. (We will quantitatively justify the reduction in the current-carrying fraction in section V, in which we describe a random resistor network simulation which proves that the majority of the current is carried by a small fraction of isolated filaments.)

To estimate the reduction in $d_F$, we use our experimentally-estimated $T_J = 1.71$ K to extract a Josephson energy[38] $E_J \equiv 2k_B T_J/\pi = 94$ μeV, and hence an inter-filamentary critical current $I_c^f \equiv 2eE_J/\hbar = 46$ nA. Defining a total critical current $I_c(T = 1.71K) = 1.2$ mA from our $V(I)$ data (see Fig. 2(b) and Fig. 3(a) inset in the main text), we estimate the number of current-carrying filaments in the crystal cross-section $n_J \equiv I_c/I_c^f$ to be $2.6 \times 10^4$, considerably smaller than the $n \equiv \sim 3.5$–$2.3 \times 10^6$ filaments deduced from our LAMH fitting. The ratio of these values $n_J/n$ provides an approximate fraction of the total crystal cross-section carrying a supercurrent, thus implying a typical filamentary diameter $d_F = 0.41$–0.74 nm. The Se-Se diameter of a single $(Mo_6Se_6)_\infty$ chain is 0.60 nm: we conclude that for $T_J < T < T_{ons}$, single $(Mo_6Se_6)_\infty$ chains are behaving as 1D superconducting filaments, as expected from the electronic structure.

## V. ANISOTROPIC RANDOM RESISTOR NETWORK

As we have demonstrated in section IV, the filamentary diameter $d_F$ which we extract from our LAMH fits is only physically meaningful if the current flows inhomogeneously through a $Na_{2-\delta}Mo_6Se_6$ crystal. Here we justify this assertion of inhomogeneity by simulating the current flow through a q1D material using an anisotropic random resistor network (Fig. S5(b)). Our model consists of a 2D $m \times n$ array of nodes, each connected to its 4 nearest neighbors by a resistor. Anisotropy is incorporated by setting the transverse (inter-chain) resistance $R_{inter}$ to be a factor of $10^3$ higher than the longitudinal (filament) resistance $R_{fil}$, motivated by the reported resistivity ratio $\rho_\perp/\rho_{//} \sim 10^3$ in $Tl_2Mo_6Se_6$.[39] Current injected at the base of the array therefore principally flows vertically through $R_{fil}$. However, to simulate the effects of the Na vacancy disorder we randomly increase 10% of $R_{fil}$ by a factor of $10^9$, thus inserting "breaks" in the chain and forcing the current to take a percolative path through our q1D pseudocrystal. The current distribution is calculated by applying Kirchoff's law to each node:

$$\sum_j \sigma_{ij}(V_i - V_j) = 0 \qquad (6)$$

where $\sigma_{ij} \equiv 1/R_{ij}$ are the conductances between nodes $i$ and $j$, $V_{i,j}$ are the voltages at each node in the array, and we employ the boundary conditions $V = 1$ at the base of the array and $V = 0$ at the top. This creates a set of $m \times n$ coupled simultaneous equations, which we solve by matrix inversion. The results for a $120 \times 120$ array are shown in Fig. S5(c): it is immediately clear that the current distribution within the array is highly inhomogeneous. Linecuts normal to the current flow show that roughly 50% of the current is carried by only $\sim 10\%$ of the filaments (Fig. S5(d)), thus confirming that electrical transport is not uniform across macroscopic $Na_{2-\delta}Mo_6Se_6$ crystals. We stress that our selection of $R \to 10^9 R_{fil}$ to simulate disorder is an arbitrary choice and similar results are obtained even for much weaker disorder potentials.

A further source of inhomogeneity stems from the current injection profile. Electrical contacts to the crystal were made by sputtering Au contact pads onto the cleaned crystal surface, then attaching 50 μm Au wires using silver epoxy (see Methods). Although we attempted to thoroughly soak each end of the $Na_{2-\delta}Mo_6Se_6$ crystals in epoxy, it was not possible to ensure that the ends of the crystal were uniformly coated with Au during the preceding sputtering procedure. Even small local variations in the contact resistance due to an inhomogeneous Au coating will accentuate the intrinsic inhomogeneity in the current flow due to one-dimensionality and disorder, thus leading to the highly inhomogeneous current distribution which we infer from our measurements.

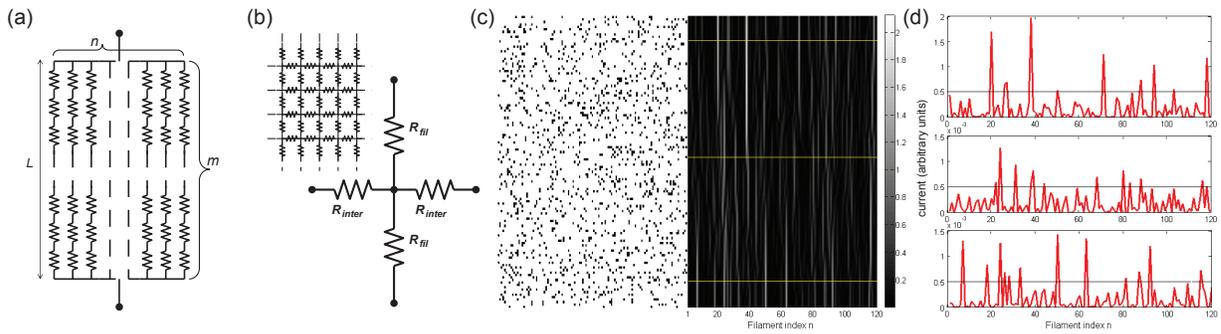

Figure S 5: Resistor networks for modelling the superconducting transitions of $Na_{2-\delta}Mo_6Se_6$. (a) Uniform $m \times n$ array of identical resistors, used to estimate an upper limit for the superconducting filament diameter from the LAMH fits to $R(T)$. (b) Random resistor network for simulation of the current flow through a q1D filamentary material. $R_{inter} = 1000$ is the transverse (inter-filamentary) resistance, while $R_{fil}$ is the longitudinal (filamentary) resistance. 90% of the longitudinal resistors $R_{fil}$ are set to 1 in order to reproduce the crystal anisotropy, while a randomly-chosen 10% of $R_{fil}$ are set to $10^9$ to accommodate the effects of Na disorder. (c) A typical random Na vacancy distribution in a $120 \times 120$ array (left) and the resultant current distribution calculated using equation 7 (right). (d) Cross-sectional current distributions at the three yellow linecuts in (c). Peaks exceeding 0.5 (black horizontal lines) are used to quantify the disproportionately large current carried by a small fraction of the filaments.

## VI. WERTHAMER-HELFAND-HOHENBERG (WHH) FITTING FOR $H_{c2}(T)$

We estimate the zero-temperature upper critical fields $H_{c2\perp,//}$ using the Werthamer-Helfand-Hohenberg (WHH) model:[40]

$$\ln \frac{1}{t} = \sum_{n=-\infty}^{\infty} \frac{1}{|2n+1|} - \left[|2n+1| + \frac{\overline{h}}{t} + \frac{(\alpha \overline{h}/t)^2}{|2n+1| + (\overline{h} + \lambda_{so})/t}\right]^{-1} \quad (7)$$

where $t = T/T_c$, the Maki parameter $\alpha = -5.2758 \times 10^{-5} \left.\frac{dH_{c2}}{dT}\right|_{T_c}$, the reduced magnetic field $\overline{h} = -(4/\pi^2)H_{c2}/\left.\frac{dH_{c2}}{dt}\right|_{t=1}$ and $\lambda_{so}$ is the spin-orbit coupling. Our fits yield $3.66$ T $\leq H_{c2\perp}(0) \leq 5.02$ T and $15.7$ T $\leq H_{c2//}(0) \leq 18.0$ T.

Within our experimentally-accessible field range, $H_{c2}(T)$ varies approximately linearly with temperature for both field orientations and so the values of $\lambda_{so}$ required to reproduce our data are unphysically large: $\lambda_{so} = 40$ for $H_{c2\perp}$ and $17 \leq \lambda_{so} \leq 80$ for $H_{c2//}$. This suggests that orbital limiting may not be the principal factor contributing to the suppression of superconductivity. It is well known that multiband superconductors with at least one dirty band can exhibit linear $H_{c2}(T)$ behavior followed by an upturn at low temperature.[41] However, a single Mo $d_{xz}$ helix band crosses the Fermi level in $Na_{2-\delta}Mo_6Se_6$: multiband physics therefore cannot be applicable. Furthermore, the open nature of the Fermi surface in $Na_{2-\delta}Mo_6Se_6$ implies that Landau quantization of the electron orbitals does not occur in high fields, ruling out any divergent $H_{c2}$ in the quantum limit.[42] While the quasi-linear trend in $H_{c2\perp}(T)$ may be attributed to a suppression of orbital limiting due to Cooper pair confinement[43] (which shares a similar physical origin to the field-induced localization observed for large $H_\perp$), the mechanism controlling $H_{c2//}(T)$ and any eventual FFLO phase remains unknown. Further work at a high magnetic field facility will be required to track $H_{c2//}$ down to milliKelvin temperatures; in the meantime, the WHH model for dirty single-band superconductors provides our best estimate for $H_{c2//}(0)$.

## VII. JOSEPHSON ENERGY SIMULATION

In a generalized inhomogeneous superconductor, the Josephson coupling between phase-disparate superconducting islands can be expressed by:

$$E_J = \frac{\pi \hbar}{4e^2 R_T} \triangle(T, H, I) \tanh\left(\frac{\triangle(T, H, I)}{2k_B T}\right) \quad (8)$$

Here, $R_T = E_a/e^2\omega_T$ is the tunneling resistance between superconducting filaments with a characteristic tunnelling frequency $\omega_T$:

$$\omega_T = A\exp\left[-\left(\frac{2m_e s^2(\phi-eV)}{\hbar^2}\right)^{1/2}\right]\cdot\exp\left[-\frac{E_a}{k_B T}\right] \quad (9)$$

where $s$ is the barrier width, $(\phi-eV)$ is the effective barrier height for tunnelling between filaments and $A$ is a constant.[44] We do not consider any periodic Fraunhofer oscillations in $E_J$ within a magnetic field,[45] since these will be suppressed due to variations in size and homogeneity across the thousands of Josephson junctions in our macroscopic crystals.

In granular superconductors (for which these equations were originally derived), $E_a$ is a characteristic energy scale describing the quantized level mismatch and typical capacitive charging energy $\frac{2e^2}{C}$ of each superconducting grain.[46,47] In $Na_{2-\delta}Mo_6Se_6$ we are not dealing with charged nanograins, but rather with conducting nanofilaments which are strongly influenced by Na vacancy disorder. Fig. S6(a) illustrates the effect of this disorder on the electrical transport: a minimum in $R(T)$ at $T_{min} \sim 30$ K is followed by a divergence at lower temperature, prior to the onset of superconductivity. This divergence is accurately described by a variable range hopping (VRH) model[48]:

$$R \propto \exp\left(\frac{T_0}{T}\right)^{\frac{1}{1+d}} \quad (10)$$

using fitting parameters $T_0 = 71\pm 5$ K and dimensionality $d = 1.4\pm 0.05$. $d$ was deliberately left as a free parameter during fitting and the resultant value of 1.4 corresponds closely to the calculated VRH exponent $\nu \equiv 1/(1+d) = 0.4$ for q1D electron crystals.[49] Importantly, $d > 1$, implying that the Coulomb repulsion is screened or weak.[50]

Transport via variable range hopping is a characteristic of strongly localized electrons. Given the excellent structural quality of our crystals as determined from our XRD analysis, this localization must be a direct consequence of the Na vacancy disorder. The magnetic field dependence of the VRH temperature $T_0(H)$ may be estimated from the normal-state magnetoresistance $R(H) \propto \exp\left(\frac{T_0(H)}{T}\right)^{\frac{1}{1+d}}$. $R(H)$ at $T = 1.8$ K (below the onset of superconducting fluctuations) is plotted in Fig. S6(b): the high-field magnetoresistance is strongly negative as expected for strongly-localized electrons,[51] approximately independent of field orientation and can be fitted by an exponential decay $a + b\exp(-cH)$ to obtain a numerical expression for $T_0(H)$.

In the absence of any capacitive effects in $Na_{2-\delta}Mo_6Se_6$, we use the VRH activation energy

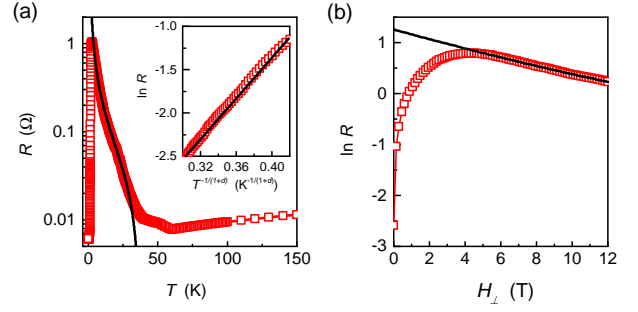

Figure S 6: Normal-state electrical transport in $Na_{2-\delta}Mo_6Se_6$. (a) Zero-field electrical resistance in $Na_{2-\delta}Mo_6Se_6$ (red symbols), illustrating a minimum at $T_{min} \sim 60$ K and a subsequent divergence prior to the superconducting transition. At low temperature, $R(T)$ is well-fitted by a VRH model (black line, see text for details). Inset: rescaled raw data $\ln R$ vs. $T^{-1/(1+d)}$ with $d = 1.4$ (red symbols), illustrating the linear behavior characteristic of VRH (black line). (b) Magnetoresistance $\ln R(H)$ at 1.8 K (red symbols). The initial rise in the resistance is due to a suppression of superconducting fluctuations. At higher fields, a strong negative magnetoresistance is observed which decays exponentially (black line).

$E_a(T,H)$ to evaluate $E_J$ in equations 8,9:

$$E_a(T,H) \equiv \frac{\partial \ln R}{\partial(k_B T)^{-1}} = \frac{k_B T_0(H)}{1+d}\left(\frac{T_0(H)}{T}\right)^{\frac{1}{1+d}-1} \quad (11)$$

To achieve the reentrant superconductivity which we observe as a function of temperature, current and magnetic field, peaks must develop in $E_J(T), E_J(H), E_J(I)$ for non-zero $T, H, I$ respectively. We consider these cases individually below.

**Temperature**

To evaluate $E_J(T)$ we require the temperature dependence of the superconducting gap $\triangle(T)$ and the activation energy $E_a(T)$ for electron transport between filaments. $\triangle(T)$ is obtained from a numerical solution of the s-wave BCS gap equation:

$$\triangle_k(T) = -\frac{1}{2}\sum_{k'} V^{(0)}_{kk'}\frac{\triangle_{k'}}{\sqrt{|\varepsilon^2_{k'}|+|\triangle^2_{k'}|}}\tanh\frac{\sqrt{|\varepsilon^2_{k'}|+|\triangle^2_{k'}|}}{2T} \quad (12)$$

where $V_{kk'}$ is the pairing potential and $\sqrt{|\varepsilon^2_{k'}|+|\triangle^2_{k'}|}$ is the quasiparticle excitation energy. $E_a(T)$ is evaluated using our experimentally-determined VRH parameters $T_0$ and $d$. We plot $\triangle(T)$ and $R_T^{-1}(T)$ in Fig. S7(a); the upper and lower limits to $E_J(T)$ shown in Fig. 4(a) of the main text are calculated using the errors in $T_0$ and $d$ from our VRH fitting procedure.

**Field**

The reentrant behavior which we observe in a magnetic field is perhaps the least obvious to predict, since at first glance there is no field-dependent term in equation 1

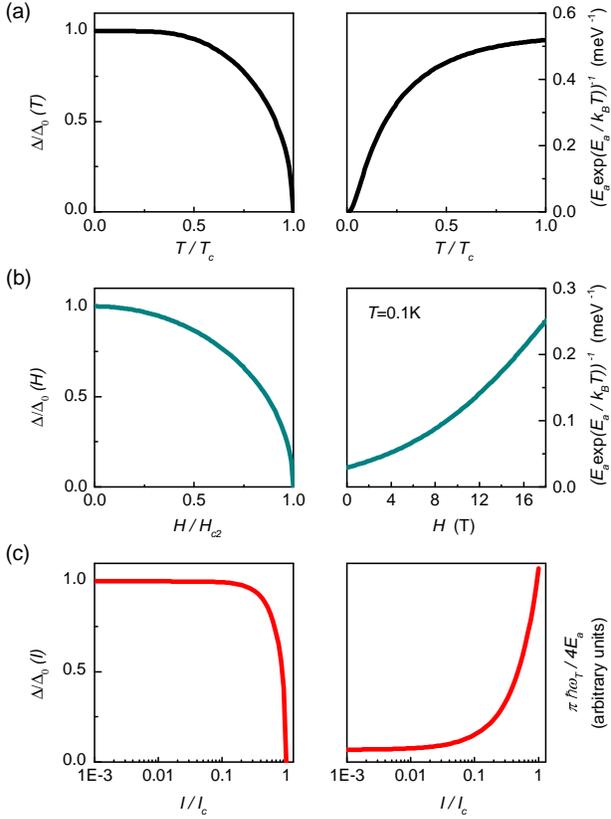

Figure S 7: Origins of the peak in the Josephson energy at high $(T, H, I)$.
(a) Factors controlling $E_J(T)$. Left: temperature-dependent BCS gap $\triangle(T)$. Right: $(E_a(T)\exp(E_a(T)/k_BT))^{-1}$, which is proportional to the inter-filamentary tunnelling conductance $R_T^{-1}(T)$. (b) Factors controlling $E_J(H)$. Left: field-dependent BCS gap $\triangle(H)$. Right: $(E_a(H)\exp(E_a(H)/k_BT))^{-1}$, which is proportional to $R_T^{-1}(H)$. (c) Factors controlling $E_J(I)$. Left: current-dependent BCS gap $\triangle(I)$. Right: $\pi\hbar\omega_T(I)/4E_a$, which is proportional to $R_T^{-1}(I)$.

which will increase $E_J$ for non-zero field. However, we recall that strongly-localized electrons exhibit an exponentially large negative magnetoresistance. Within the VRH transport model, this can be expressed as a field-dependent localization temperature:

$$\rho(H) = \rho_0 \exp\left(\frac{T_0(H)}{T}\right)^{\frac{1}{1+d}} \quad (13)$$

We use our $\rho(H)$ curve at $T = 1.8$ K (Fig. S6(b)) to determine $T_0(H)$. Once the superconductivity and fluctuation paraconductivity have been suppressed ($H > 5.2$ T), we fit $\rho(H)$ using the expected exponential decay $a + b\exp(-cH)$. We now have:

$$a + b\exp(-cH) = \rho_0 \exp\left(\frac{T_0(H)}{T}\right)^{\frac{1}{1+d}} \quad (14)$$

which we may rearrange and solve for $T_0(H)$. Substituting our numerical expression for $T_0(H)$ into equation 11, we obtain a field-dependent activation energy $E_a(H)$ which balances $\triangle(H)$, the field dependence of the superconducting gap. $\triangle(H)$ can be approximated as[52]:

$$\triangle(H) = \triangle_0 \left(1 - \left(\frac{H}{H_{c2}}\right)^2\right)^{\frac{1}{2}} \quad (15)$$

in which $H_{c2}$ is determined from our WHH fits (Fig. 3(f) in the main text). $\triangle(H)$ and $R_T^{-1}(H)$ are plotted in Fig. S7(b); the two $E_J(H)$ curves for each field orientation in Fig. 4(a) of the main text are calculated using the upper and lower WHH limits for $H_{c2//,\perp}$.

**Current**

In a transport experiment, a voltage $V$ must be applied for a current $I$ to flow. For our crystals, the normal-state resistance is approximately Ohmic, i.e. $V = IR_{\text{NS}}$ where $R \approx 1\Omega$. This is justified by the linear $I-V$ curves which we acquire in the normal state (Fig. 2(b) in the main text; the slight non-linearity at high current is principally due to sample heating effects). A peak in $E_J$ is formed by balancing $\exp\left[-\left(\frac{2m_es^2(\phi-eV)}{\hbar^2}\right)^{1/2}\right]$ with $\triangle(I)\tanh\left(\frac{\triangle(I)}{2k_BT}\right)$: since we have $V \sim I$, we may substitute $I$ for $V$ in order to compare these two expressions.

Since we know neither the barrier height $\phi$ nor width $s$, calculating $E_J(I)$ is daunting and we are obliged to approximate heavily: for this reason, we do not provide any estimate of the error in $E_J(I)$. We start by exploiting the apparent analogy between the onset of phase coherence in our q1D crystals with 2D materials exhibiting BKT transitions, where $E_J = \frac{2}{\pi}k_BT_{BKT}$ at $T_{BKT}$. Setting $T_J \equiv T_{BKT} = 1.71$ K, we evaluate $\triangle(T = T_J)$ and $E_a(T = T_J)$ using $T_0$ and $d$ from our VRH fits, then substitute these numerical values into equation 8 (setting $A=1$), leaving us with $\frac{2m_es^2(\phi-I)}{\hbar^2} = 775$. Now we are finally forced to make an educated guess at the average barrier width: here we cannot simply assume that this is equivalent to the interchain separation $\sim 0.64$ nm, since we must take into consideration the real-space decay of the amplitude of the localized electron wavefunctions. The *effective* barrier width therefore ranges from 0.64 nm up to the 1D localization length[53] $\xi_L \equiv 4/k_BT_0N_{EF}^{1D} = 380$ nm (where $N_{EF} = 1.07 \times 10^{28}$ J$^{-1}$m$^{-1}$ is the 1D density of states, which we estimate using the experimentally-determined 0.055 states eV$^{-1}$ atom$^{-1}$ for In$_2$Mo$_6$Se$_6$).[3] The tunnelling probability decays exponentially with barrier width and so we opt for the logarithmic average of these values, 16 nm, which yields a physically reasonable effective barrier $\phi - eV = 0.1155$ eV. Since the $T_J$ and $T_{ons}$ which we use in the calculation of $\phi - eV$ were acquired at $I = 6\times 10^{-4}$ A and we know that $I \sim V$, we set $\phi = 0.1161$ eV in our simulation of $E_J(I)$.

The current dependence of the superconducting gap $\triangle(I)$ is still an open question theoretically. However,

we may derive an approximate relation using Ginzburg-Landau theory:

$$\frac{|\psi|^2}{\psi_\infty^2} = 1 - \left(\frac{\xi m_e v_s}{\hbar}\right)^2 \quad (16)$$

where $\psi$ is the Ginzburg-Landau order parameter and $v_s$ the superfluid velocity. We recall that the supercurrent density $J_s = 2e n_s v_s$, where the superfluid density $n_s = |\psi|^2$, i.e. $J_s$ varies linearly with $v_s$ (except close to $J_c$). Now, $\frac{|\psi|^2}{\psi_\infty^2} \propto \left(\frac{\triangle(I)}{\triangle_0}\right)^2$; to impose gap closure at $I_c$ we therefore approximate $\triangle(I)$ using

$$\triangle(I) = \triangle_0 \left(1 - \left(\frac{I}{I_c}\right)^2\right)^{\frac{1}{2}} \quad (17)$$

The similarity of this relation to the gap variation in a low-dimensional superconductor in a magnetic field (equation 15 above) should not be surprising, since if we disregard vortex penetration then the current-induced uniform magnetic field varies linearly with the applied current $I$. Very close to $I_c$, equation 17 breaks down and the gap is expected to close more rapidly; however this occurs too late to have any effect on the peak formation in $E_J(I)$. We plot the resulting $\triangle(I)$ and $R_T^{-1}(I)$ in Fig. S7(c). Despite the numerous approximations required to estimate $E_J(I)$, the correspondence between the maximum in $E_J$ and the resistive minima (Fig. 4(a,b) in the main text) is strikingly accurate, thus validating our reasoning.

## VIII. TUNING REENTRANCE BY VARYING THE DISORDER LEVEL

Until now, we have focused on tuning the reentrant phase coherence by modulating $(T, H, I)$, since these are the key parameters which will vary during practical applications of a nanofilamentary superconductor. It is also instructive to consider the effect of the disorder level on the reentrance, since future methods of synthesizing nanofilamentary superconductors are likely to enable some tuning of the eventual disorder.

**Reentrance in a less-disordered crystal**
The Na vacancies responsible for the disorder in $Na_{2-\delta}Mo_6Se_6$ form naturally due to the high crystal growth temperature (1750°C) and we cannot precisely control their concentration. However, we may infer the resultant disorder level in our crystals using two parameters: the resistivity at room temperature $\rho(300K)$ and the variable range hopping temperature $T_0$ which describes the normal-state divergence of $\rho(T)$ at low temperature. $\rho(300K)$ and $T_0$ both rise with disorder. A detailed study of the evolution of the normal-state $\rho(T)$ and $T_{ons}, T_J$ with disorder is beyond the scope of this work and will be published elsewhere.[54] Here, we restrict ourselves to examining reentrance at low temperature in a second crystal with reduced disorder compared with that studied in our manuscript: $\rho(300K) = 9.9 \times 10^{-7}$ $\Omega$m, $1.1 \times 10^{-6}$ $\Omega$m and $T_0 = 112$ K, 180 K respectively. Figure S8(a) displays $R(T)$ curves acquired in this less-disordered crystal at a range of applied currents. The similarities with the data in Figs. 2,3(a) in the main text are immediately apparent: an upturn in $R(T)$ emerges at low temperature as the current is reduced, indicative of reentrant phase coherence. A familiar finite-resistance plateau develops for $I < 10\mu A$, while the onset of transverse phase coherence is supported by BKT-style exponential scaling in $R(T)$. Reentrance is also observed for increasing current as $T \to 0$ (Fig. S8(b)), just as seen in Fig. 4(a) from the main text. However, the reentrance threshold $T_R$ (below which $R(T)$ rises steeply) lies at much lower temperature: $\sim 0.45$ K, compared with $\sim 1.1$ K in the more disordered crystal (Fig. S8(c)). This is consistent with the concept of localization-driven reentrance: less disorder implies that a lower temperature will be required to reduce the Cooper pair wavefunction overlap sufficiently strongly to decouple the filaments.

We therefore believe that reentrant phase coherence can persist in nanofilamentary superconductors as long as there is sufficient disorder to cause localization in the normal state (i.e. $dR/dT < 0$, $dR/dH < 0$). The reentrance threshold $T_R(H, I)$ will fall to zero as the disorder is reduced. In the weak disorder limit (where the normal state remains metallic), we expect behaviour simi-

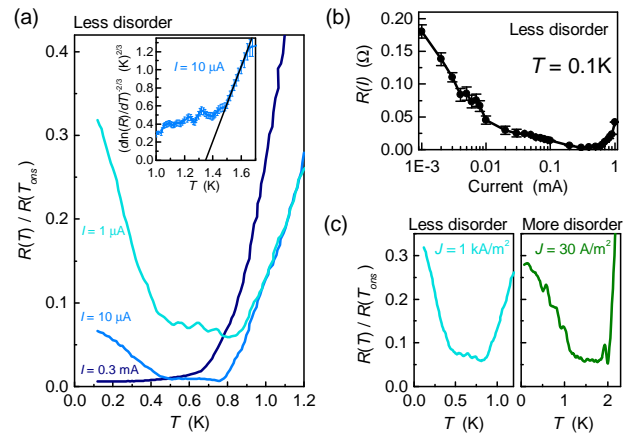

Figure S 8: Electrical transport data from a less-disordered $Na_{2-\delta}Mo_6Se_6$ single crystal.
(a) $R(T)$ curves at three different currents, illustrating reentrant phase coherence for $(T, I) > 0$. Inset: BKT-type exponential scaling (similar to Fig. S4(e)), indicating $T_J \sim 1.35$ K with a characteristic hump forming around/below this temperature. [Note that $T_{ons} \sim 2.0$ K in this crystal.] (b) $R(I)$ acquired at 0.1 K in the less-disordered crystal, illustrating current-induced reentrance even as $T \to 0$. (c) Comparison of the finite-resistance plateau in less-disordered (left) and more-disordered (right) samples for $I = 1\mu A$. No peak is visible in $R(T)$ in the less-disordered crystal since $T_J$ lies above the $T$-axis limit and the current density is higher due to the smaller crystal diameter.

lar to that seen in Tl$_2$Mo$_6$Se$_6$, in which a 1D→q3D dimensional crossover occurs,[11] but no reentrance is observed.[3,39] Conversely, for extremely disordered samples, we do not expect to observe any reentrance. Although the onset of superconducting fluctuations will remain visible in $R(T)$ as a peak, $R(T)$ will remain large at all temperatures due to a combination of emergent spatial inhomogeneity in the order parameter and the increasing prevalence of quantum phase slips. This situation will be discussed in detail elsewhere.[54] A finite yet easily-accessible disorder window therefore exists for developing nanofilamentary superconductors with reentrant phase coherence.

### Deducing the origins of the finite-resistance plateau

Data from this less-disordered crystal are also helpful to understand the origin of the finite-resistance plateau which forms in both crystals for $I < 10\mu$A. In Fig. S8(c) we compare $R(T)$ curves acquired at $I = 1\mu$A in both crystals. Given that non-zero resistance in a superconductor is a signature of inhomogeneity or disorder, one might therefore expect the finite-resistance plateau to be reduced or absent in the less-disordered crystal. On the contrary, the plateau remains prominent and may even be enhanced: at 1 $\mu$A the plateau lies at a similar height ($\sim 7\%$ of $R(T_{ons})$) to that in the more disordered crystal, despite the fact that the current density $J$ in the less disordered crystal is much higher (1 kA m$^{-2}$ vs. 30 A m$^{-2}$) - and increasing the current clearly suppresses the plateau (Figs. 2,3(a) in the main text, Fig. S8(a)). We therefore find no obvious link between the plateau and the intrinsic (Na vacancy) disorder responsible for localization.

Instead, we believe that this plateau originates from isolated cracks, twin boundaries or other large-scale extrinsic barriers which separate crystalline regions exhibiting transverse phase-coherence, i.e. long-range superconducting order. This is supported by the plateau resistance remaining approximately constant over a broad temperature range ($> 1$ K for $I = 0.5\mu$A in Fig. 3(a) from the main text). If no long-range order existed within this region, $R(T)$ would rise steeply as temperature falls, since in this case current would only pass between filaments via quasiparticle tunneling (and the quasiparticle density is rapidly suppressed by the opening of the superconducting gap). The flat plateau is incompatible with such quasiparticle transport and instead implies that the majority of the crystal is phase-coherent. Together with the systematic appearance of a peak/hump feature (SI section IIID) and the fact that our experimental data are well-described by a Josephson coupling model which assumes transverse phase coherence (Fig. 4(b) in the main text), this provides compelling evidence for the presence of long-range order within the plateau regime, despite its non-zero resistance. Raising the current increases the electron tunneling frequency across the extrinsic micro-cracks/barriers responsible for plateau formation: the resultant decrease in the barrier resistance eventually enables supercurrents to cross the barriers by Josephson tunneling, suppressing the plateau for $I \geq 10\mu$A. However, the transverse phase coherence remains reentrant even at larger currents (up to $\sim 0.2$ mA in Fig. S3, for example). This confirms the distinct energy scales of the plateau and the reentrance. We therefore attribute the plateau to isolated extrinsic defects within the crystals, and the reentrance to intrinsic Na vacancy-induced localization.

## IX. ADDITIONAL REENTRANCE MECHANISMS IN Na$_{2-\delta}$Mo$_6$Se$_6$

Regarding the possible role of other (non-Josephson) effects in Na$_{2-\delta}$Mo$_6$Se$_6$, we note that the observation of reentrance in zero magnetic field (Figs. 2(a), 3(a), 4(a,b)) absolves exchange-field compensation, geometric vortex pinning,[55] Fermi surface reconstruction[56] and Fulde-Ferrell-Larkin-Ovchinnikov (FFLO) phase formation[43,57,58] from responsibility for this phenomenon. Magnetic impurities have also been predicted[59] and observed[60] to enhance superconductivity in a magnetic field, even causing reentrance for a narrow $(T, H, I)$ parameter range. However, this mechanism is unlikely to play an important role in Na$_{2-\delta}$Mo$_6$Se$_6$, since $M_2$Mo$_6$Se$_6$ crystals show no evidence for any significant magnetic impurity content.[3] Furthermore, the reentrant phase coherence which we report in Na$_{2-\delta}$Mo$_6$Se$_6$ is visible at zero field and in small currents $\ll I_c$, in contrast to magnetic impurity-induced reentrance which only develops for currents close to $I_c$ in a non-zero magnetic field.

However, we do not exclude the presence of a FFLO phase at high fields in Na$_{2-\delta}$Mo$_6$Se$_6$. The weak-limit BCS Pauli limit $H_P \equiv 1.84T_{ons} = 5.0$T: we observe reentrant phase coherence at $H_{//} = 8.5$ T, considerably above $H_P$ (Fig. 4(a) in the main text). Furthermore, the WHH-estimated $H_{c2//} \sim 16$-18 T (Fig. 3(f) in the main text; section VI) is more than three times larger than $H_P$ and seems far too high to be explained by spin-orbit scattering alone. If the Maki parameter $\alpha \equiv \sqrt{2}H_{c2//}/H_P > 1.6$, a FFLO phase may form at temperatures below 0.55 $T_{ons}$.[61] Our large Maki parameter $4.4 \leq \alpha \leq 5.1$ therefore encourages a FFLO scenario at low temperature in Na$_{2-\delta}$Mo$_6$Se$_6$. Although FFLO phases are generally incompatible with disorder, this condition may be lifted if the nodes in the spatially-modulated FFLO order parameter coincide with crystal defects. The evolution of $H_{c2//}(T)$ in Na$_{2-\delta}$Mo$_6$Se$_6$ clearly merits future experimental attention.

\clearpage

\end{document}